\documentclass{article} 

\usepackage{cite}
\usepackage{times}
\usepackage{hyperref}
\usepackage{url}
\usepackage{amssymb}
\usepackage{subcaption}
\usepackage{graphicx}
\usepackage{float}
\usepackage{bbm}
\usepackage[noend]{algpseudocode}
\usepackage{amsmath}
\usepackage{amsthm}
\usepackage{hyperref}
\usepackage{enumitem}
\usepackage[]{algorithm2e}
\usepackage{subcaption}
\newtheorem{prop}{Proposition}
\usepackage[usestackEOL]{stackengine}
\usepackage{titlesec}

\usepackage{float}
\usepackage[margin=1.5in]{geometry}

\title{Overcomplete Frame Thresholding for\\Acoustic Scene Analysis}

\author{Romain Cosentino, Randall Balestriero, Richard Baraniuk, Ankit Patel \\
Department of Electrical and Computer Engineering\\
Rice University\\
 6100 Main St, Houston, TX 77005 \\
}

\date{}

\begin{document}
\maketitle

\begin{abstract}
In this work, we derive a generic overcomplete frame thresholding scheme based on risk minimization. Overcomplete frames being favored for analysis tasks such as classification, regression or anomaly detection, we provide a way to leverage those optimal representations in real-world applications through the use of thresholding. We validate the method on a large scale bird activity detection task via the scattering network architecture performed by means of continuous wavelets, known for being an adequate dictionary in audio environments.
\end{abstract}

\section{Introduction}
\label{Intro}
From anomaly prediction in medical data to image recognition, the input signal is often corrupted with noise \cite{ganeshapillai2011reconstruction,danahy2005non,xie2012image}. A classical approach in signal processing is based on the following analysis-synthesis framework: the data are analyzed (change of basis), denoised in their new basis, and then either fed to a machine learning classifier or synthesized (inverse the change of basis) before being processed by another algorithm. There is a wild number of analysis-synthesis operator, each of them being well defined mathematically by their specific properties. In this work, we will focus on overcomplete frame operators. As a matter of fact, those operators have been applied and studied in a wide range of areas such as wireless communications \cite{strohmer2003optimal}, signal processing \cite{daubechies1990wavelet,goyal1998quantized}, quantum computing \cite{eldar2002optimal}, functional analysis \cite{lacey1997lp}. In fact, they yield interesting properties such as robustness \cite{casazza2012finite}, stability and compactness \cite{balan2006density}. 
A simple intuition behind those properties can be trivially described via the human vocabulary analogy. Consider the atoms of a dictionary as the vocabulary of a person. The higher the cardinal of the dictionary is, the higher the person will be precise and concise in his saying. As such, the analysis of his discourse will be eased thanks to its sparsity. In signal processing, sparse representations provide a fruitful pre-processing step allowing simple machine classifiers to produce high accuracy results. 
In this work, we will focus on overcomplete frames yielding a time-frequency representation as having an efficient representation of a signal is a synonym of extracting both its time and frequency content \cite{meyer2006,d2012wavelet}. Especially, wavelet frames are known to provide a robust time-frequency representation for non-stationary signals as it is localized, both, in time and frequency. 
In most cases, wavelets can be used for three different tasks. First, for compression purposes: in fact, it has been demonstrated that wavelet transforms provide sparse representations particularly adapted for compression of natural signals with non-stationary components \cite{Wavtour,moi}. Secondly, for denoising purposes. Due to the sparsity of the induced representation and the ability to derive analytical solutions, optimized denoising solution with reconstruction guarantees have been developed \cite{donoho1995adapting,donoho1994ideal}. Finally, the last task corresponds to analysis in which one aims at detecting subtle patterns of the input signal by using over-complete dictionaries. The analysis task can be considered today as crucial, especially when considering the need for pre-processing for current machine learning tasks. Unfortunately, thresholding methods for overcomplete frames have been tackled for special cases.

Therefore, we propose to remedy this absence by deriving a general analytical thresholding scheme for overcomplete frames principled via empirical risk minimization being computationally efficient and with the ability to leverage overcomplete frameworks within a corrupted environment. 

The paper is organized as follows: \ref{sect2} and \ref{contrib} are devoted to the related work and contribution of the paper,  the section \ref{denoise} shows the theoretical results including the derivation of the over-complete hard-thresholding technique, derivation of error bounds and presentation of computational complexity and pseudo-code. Finally, in section \ref{validation} we provide experiments demonstrating the ability of the proposed scheme via the introduction of a Sparse Deep Scattering Network (SDSN), an extension of the Deep Scattering Network (DSN) \ref{DeepCroisee}. The task comparing the DSN and SDSN consists of a bird detection task using the Freefield1010\footnote{http://machine-listening.eecs.qmul.ac.uk/bird-audio-detection-challenge/}  audio scenes dataset in \ref{validation}. The appendix is divided into two parts, Appendix \ref{appendixA} provides both, the pre-requisite and details about the wavelets families used in our experiments;  In Appendix \ref{appendixC}, we propose a review of the orthogonal denoising framework proposed by Donoho et. al. as well as the  mathematical details and proofs for the over-complete thresholding non-linearity that we introduce.

\subsection{Related Work}
\label{sect2}
During the last decades, the properties of wavelets have been exploited to perform denoising on a wide range of signals \cite{cohen2012signal,taswell2000and,rangarajan2002image}. In fact, the decomposition of a signal by mean of a wavelet transform concentrates the signal of interest in few coefficients having high energy, while the noise is uniformly propagated throughout all the time-frequency components \cite{teolis1998computational}. This property is the direct consequence of the number of vanishing moment of the considered wavelet. In fact, it is defined as their orthogonality with respect to a certain order $p$ of regular signals $ y \; \in \mathcal{C}^p$. This property makes them an edge-detector basis (in the broad sense of the term) capable to provide sparse representations.
Donoho et. al. via a statistical signal processing approach to the problem, were able to provide powerful orthogonal basis denoising techniques holding theoretical guarantees \cite{donoho1998minimax,donoho1995adapting,donoho1994ideal,donoho1995wavelet}. However, orthogonal wavelet decomposition provides poor time-frequency resolution and are more prone to instabilities.
Since overcomplete frames provide extremely precise and often sparse representations, different approaches have been taken to tackle special cases of an overcomplete dictionary. This ability to fit the signal such closely should allow for more precise and robust denoising capacities.
An extension to the work of Donoho et. al., one step ahead toward overcomplete frames, has been achieved by Berkner et. al. in \cite{berkner1998correlation} using Translation Invariant Discrete Wavelet Transform and Biorthogonal-DWT. Then, using independent Component Analysis, Marusic et. al. \cite{marusic2005image} were able to tackle denoising in the case of multiple DWT. Others are exploiting the tools available in sparse coding-based models to perform denoising in learned "overcomplete frames" \cite{zhao2013image,carrera2017sparse}. Therefore as these work rely on specific structures of overcomplete dictionaries that are often based on orthogonal filters drastically limit their applications.

\subsection{Contributions}
\label{contrib}
The contribution of the paper is the derivation of a general analytical thresholding technique for overcomplete frames. This technique based on empirical risk minimization will be studied by providing error bounds, computational complexity as well as pseudo-code allowing one to implement the method. We demonstrate the generality and robustness of our approach by mean of redundant continuous wavelet dictionaries. First, we propose the visualization of the thresholded scalograms as well as the evaluation of the sparsity before and after denoising. Secondly, we propose quantitative experiments validating the proposed scheme. Those later are achieved via introducing our thresholding technique into the Deep Scattering Network (DSN), which leads to the development of the Sparse Deep Scattering Network (SDNS). We compare the performance of the SDNS and the DSN on a bird activity detection task. As we will see, bringing denoising into this state-of-the-art pre-processing technique allows to further increase its linearisation capability and improves the accuracy of the classifier for a corrupted dataset composed of extremely diverse audio scenes. 

\section{Overcomplete Frames Denoising}
\label{denoise}
In the case of analysis-synthesis framework with an overcomplete dictionary, the first question that arises is whether the mapping from the domain to the co-domain preserves distinctness. For sake of generality, we will assume that such operator has a nontrivial kernel, thus the application is not injective. Therefore, the dual of the dictionary is achieved via the pseudo-inverse transform. Notice that, Berkner et. al. in \cite{berkner1998correlation} proposed the use of the Moore-Pensore inverse to build the reconstruction dictionary of Bi-orthogonal wavelets and TIDWT dictionaries. 
Let's assume the observed signal, denoted by $y$, is corrupted with white noise such that $y= x +\epsilon$ where $x$ is the signal of interest and $\epsilon \sim \mathcal{N}(0,\sigma^2)$.
We now denote by $W \in \mathbb{C}^{(n \times (J \times Q)) \times n}
$ the matrix composed by the the filters at each time and scale (i.e: explicit convolution matrix of the operator $\mathcal{W}$) such that $\forall x \in \mathbb{R}^n$, $\mu(x) = Wx$ is the wavelet transform of $x$. In other world, for each time position and each scale, the filters vectors $\psi_i^T \in \mathbb{C}^{n}, \: \forall i \in \{1,...,n\times(J\times Q)\}$ are the row vectors of $W$. Then, we denote by $W^{\dagger} \in  \mathbb{C}^{ n \times ((J \times Q) \times n )}$ the generalized inverse of $W$ such that  $W^{\dagger}W=I$. Each column of $W^{\dagger}$, are the inverse filters $\psi^{\dagger}_i \in \mathbb{C}^{n}, \: \forall i \in \{1,...,n\times(J\times Q)\}$. The estimate of $x$ is given by  $\hat{x}_{W,D}(y) = W^{\dagger}D^{S}Wy$ where $D^{S}$ is a diagonal binary operator such that,
 \begin{equation}
 D^{S}_{i,i}= \delta_i = \left\{\begin{matrix}
 1  \: \: if \: \: i \in S  \\ 
 0  \: \: if \: \: i \in U 
\end{matrix}\right.
\end{equation}
with $U$ and $S$ denoting respectively the set of selected and unselected wavelet coefficients. We also define $D^{U}$ such that $I = D^{U} + D^{S}$. Our thesholding estimate is based on the mean-square error between the signal of interest $x$ and the synthesized thresholded observed signal $y$: 
\begin{align}
 \mathcal{R}^{\star}(x,W)&=  \min_{\delta} \mathbb{E} \left \| x-\hat{x}_{W,D}(y) \right \|^2 =  \min_{\delta} \mathbb{E} \left \| W^{\dagger} (Wx-D^{S}Wy \right \|^2, 
 \end{align}
 which can be rewritten in the following explicit form,
\begin{align}
 \mathcal{R}^{\star}(x,W) = & \min_{\delta}  \sum_{i,j=1}^{n \times (J \times Q)} \mu_{i}\mu_{j} \psi^{\dagger^{T}}_{i} \psi^{\dagger}_{j} 1_{\left \{ \delta_i = 0, \delta_j = 0 \right \} } \nonumber \\
 & + \sigma^{2} \sum_{i,j=1}^{n \times (J \times Q)}  ( \psi^{\dagger^{T}}_{i} \psi^{\dagger}_{j} ) \psi_i^{T} \psi_j 1_{\left \{ \delta_i =1,\delta_j =1 \right \}},
 \end{align}
where $\mu =  Wx$ are the co-domain's coefficients. It is clear that the correlation inherent to the redundant information contained in the dictionary implies an intractable factorial optimization problem as the ideal risk is now dependent on all the possible pairs in the frequency axis. Since this ideal risk requires the intervention of an oracle, we develop an upper bound to the ideal risk such that we benefit an explicit equation for the thresholding operator that is adapted to any over-complete transformation and possesses a tractable and analytic expression. We propose to use a min-max formulation as an upper bound of this ideal risk explicitly derived in Appendix \ref{upper-bound}. 
The upper-bound on the optimal risk is denoted by $\mathcal{R}_{up}$ and defined as,
\begin{align}
    \mathcal{R}_{up}(x,W) = \sum_{k =1}^{n(J*Q+1)}  \min(\mathcal{R}_{up}^{U}(x)[k],\mathcal{R}_{up}^{S}[k]),
\end{align}
where we denote by $\mathcal{R}_{up}^{U}$ the upper bound error term corresponding to unselected coefficients:
\begin{equation}
    \mathcal{R}_{up}^{U}(x)[k] =\sum_{j =1}^{n \times (J \times Q)} \left | \mu_{k}(x)\mu_{j}(x)    \psi^{\dagger^T}_{k} \psi^{\dagger}_{j}   \right |,
\end{equation}
which means that if we do not select a coefficient (e.g: $\mu_k$), the risk is equal to a weighted value of this coefficient. The weight are all the coefficient ($\mu_j, \forall j$) weighted by the correlation between the reconstruction filter  ($\psi^{\dagger}_k$) and each of the reconstruction filters ($\psi^{\dagger}_j , \forall j$). 
Now, let's $\mathcal{R}_{up}^{S}[k]$ be the upper bound error term corresponding to the selected coefficients:
\begin{equation}
    \mathcal{R}_{up}^{S}[k] = \sigma^{2} \sum_{j=1}^{n \times (J \times Q)}   \left |  (\psi^{\dagger^T}_{k} \psi^{\dagger}_{j}) \psi_k^{T} \psi_j  \right |.
\end{equation}
In the case where the coefficient is selected, the error corresponds to the energy of the noise propagated by the redundant frame. As a matter of fact, the noise is weighted by the correlation of, both the filter $\psi^{T}_k$ and the reconstruction filters $\psi^{\dagger}_k$, with all filters $\psi_j^T$ and reconstruction filters $\psi^{\dagger}_j$, $\forall j$.
Now, one way to evaluate this upper-bound, is to assume an orthogonal basis, and to compare it with the optimal risk in the orthogonal case which leads to the following proposition.
\begin{prop}
Assuming orthogonal filter matrix $W_O$, the upper bound ideal risk coincides with the orthogonal ideal risk:
\begin{equation*}
    \mathcal{R}_{up}(x,W_O) = \mathcal{R}_{O}(x,W_O).
\end{equation*}
\end{prop}
The proof is derived in \ref{upper-bound-comparison}. Therefore, even with the upper-bound, our optimal risk, in the case of bijective transform, can still be used and the optimal solution can be recovered. 
In order to apply the ideal risk derive, ones needs an oracle decision regarding the signal of interest. In real application, the signal of interest $x$ is unknown. We thus propose the following empirical risk:
\begin{align}
    \tilde{\mathcal{R}}(y,W) =  \sum_{k =1}^{n \times (J \times Q)}  \min(\mathcal{R}_{up}^{U}(y)[k],\mathcal{R}_{up}^{S}[k]).
\end{align}
This risk corresponds to the empirical version of the ideal risk where the observed signal $y$ is evaluate in the left part of the minimization function. In order to compare this empirical risk with the ideal version, we propose the bound analysis with respect to the optimal risks in the two possible extreme cases:
\begin{prop}
In the case where $D^S=I$, the empirical risk coincides with the upper bound ideal risk:
\begin{equation*}
   \tilde{\mathcal{R}}(y,W) =   \mathcal{R}_{up}(x,W).
\end{equation*}
\end{prop}
\begin{prop}
\label{prop:up}
In the case where $D^U=I$, the following bound shows the distance between the empirical risk and the upper bound ideal risk:
\begin{equation}
   \tilde{\mathcal{R}}(y,W) \leq \mathcal{R}_{up}(x,W) + \sum_{k,j=1}^{n \times (J \times Q)} C[k,j] \times \left |  \psi^{\dagger^T}_{k} \psi^{\dagger}_{j}   \right |, a.s.
\end{equation}
where, 
\begin{equation*}
    C[k,j] =     \left | \mu_{k}(x) \right | \left \| \psi_j \right \|_1 \sigma \sqrt{\frac{2}{\pi}} +  \left | \mu_{j}(x) \right |\left \| \psi_k \right \|_1 \sigma \sqrt{\frac{2}{\pi}}  + \sigma^2 (1- \frac{2}{\pi}).
\end{equation*}
\end{prop}
Refer to \ref{upper-bound-empirical} for proofs.
As the empirical risk introduces the noise in the left part of the risk expression, this term represents the propagation of this noise throughout the decomposition. Naturally, we from the proposition \ref{prop:up}, that the more redundant the dictionary is, the less the bound is tight. 

\subsection{Algorithm and Complexity}
In this section we present practical guide and discussion concerning the application of the method for large scale tasks.
In term of complexity, by using the proposed local method, we are asymptotically linear in time hence number of signals to treat and quadratic in the number of filters present in the over-complete filter-bank. However, the latter can be ''pushed out'' of the computations by performing it once for all signals. In fact, most of the costly terms to compute are constant for all treated signals (the pseudo inverse of the filters correlation). As a result the most demanding operation remains in the computation of the correlation matrix to be done for each column of the scalogram. Again this is quadratic in the number of filters but linear in time and with computations completely independent from one another allowing very efficient parallelization or map-reduce schemes. 
\begin{algorithm}[h]
\SetAlgoLined
\KwResult{Denoised Signals}
W$:=\mathcal{W}[\psi_0^{(\ell)},\lambda^{(\ell)}]$\\
corrW=$(W^T+\epsilon) (W+\epsilon)$\hspace{1cm}\textit{Add small noise for round-off errors}\\
M$=pinv(W)$\hspace{1cm}\textit{Compute Moore-Pseudo Inverse }\\
corrM=$M^T M$\\
 \For{ each signal $y$}{
 $U[y]=y\star W$\\
 \For{For each window $p =0,\dots,P-1$}{
 $I_p=[p\_{size}p:(p\_{size}+1)p]$\hspace{1cm}\textit{Generate indices of the corresponding window}\\
$\hat{\sigma}_p=std(U[y]_{.,I_p})/0.67$\hspace{1cm}\textit{Perform noise level estimation (MAD in this case)}\\
\For{each column $t\in I_p$ of the window}{
$a=\sum_j\left(|U[y]_{.,t}U[y]_{.,t}^T\cdot \text{corrM}|\right)_{i,j}$,\\
$b=\hat{\sigma}_p\sum_j|\left(|\text{corrM}\cdot \text{corrM}|\right)_{i,j}$,\\
mask$_{i,t} = \arg\min\left(a,b\right)$\hspace{1cm}\textit{Returns $0$ or $1$, the index of the $\min$}
}
}
$\mathcal{T}U[y]=$mask$\odot U[y]$
}
 \caption{Algorithm to compute Non-Orthogonal Denoising. Practical implementation does not use the internal for loops but a vectorize version of this algorithm for efficieny. $P$ is the total number of windows, $p\_size$ its size.}
\end{algorithm}

\section{Experimental Validation with Continuous Wavelet Transforms:\\Bird Activity Detection Task}\label{validation}
We propose to validate the two contributions over a large-scale audio dataset. 
The data set consists of $7000$ field recording signals from  freefield1010\footnote{http://machine-listening.eecs.qmul.ac.uk/bird-audio-detection-challenge/} collected via the Freesound\footnote{https://arxiv.org/abs/1309.5275} project. This collection represent a wide range of audio scenes such as birdsong, city, nature, people, train, voice, water...
The focus in this paper is the bird audio detection task that can be formally defined as a binary classification task, where each label corresponds to the presence or absence of birds. Each signal is $10$sec. long, and has been sampled at $44.1$Khz. The evaluation of the results is performed via the Area Under Curve metric on $33\%$ of the data. The experiments are repeated 50 times.
The total audio length of this dataset is thus of slightly less than $20$ hours of audio recordings. To put in comparison, it is about $20 \times$ larger than CIFAR10 in term of number of scalar values in the dataset. The results comparing our algorithm to the DSN are achieved with two different wavelet and are provided in Table \ref{table:compare}.  For all the experiments, the octave and quality parameters of the layers are $J1=5,Q1=8,J2=4,Q2=1$. As the feature of interests are birds songs, only high frequency content requires high resolution, the thresholding is applied per window of $2^{16}$ representing $\approx 1.5 sec$.
\subsection{Continuous Wavelet Transform}
We first denote the collection of scaling factors by, 
\begin{equation}
\lambda = \left \{ \lambda_j, j= 1,...,J \times Q \right \},\;\; \lambda_j=2^{(j-1)/Q},
\end{equation} 
The hyper-parameters are, $J$ and $Q$ which represents respectively the number of octave to decompose and  the quality coefficients a.k.a the number of wavelets per octave.
Based on the collection $\Lambda$, the wavelet filter-bank can be derived by scaling the mother wavelet $\psi_0$ leading to the filter-bank denoted as
\begin{align}
\mathcal{W}[\psi_0,\lambda]=\begin{pmatrix}
\psi_{\lambda_1}\\
\dots \\
\psi_{\lambda_{J \times Q}}\\
\end{pmatrix},\;\;\psi_{\lambda_j}(t)=\frac{1}{\sqrt{\lambda_j}}\psi \left(\frac{t}{\lambda_j}\right),
\end{align}
where we omit the upper-script over the $\lambda$ parameter to avoid redundancy. Also, we denote this filter-bank as $\mathcal{W}[\psi_0,\lambda]:=\mathcal{W}$. 
We define the continuous wavelet transform as,
\begin{align}
y \star \mathcal{W}= \begin{pmatrix}
y \star \mathcal{W}(1,.) \\
\dots \\
y \star \mathcal{W}(JQ,.) 
\end{pmatrix}.
\end{align}

\subsection{Qualitative Analysis}
we first provide some examples of denoising over different audio scenes.
In the Fig.\ \ref{fig:sparse}, we present the visualization and evaluation of the sparsity within the wavelet domain for several signals. The evaluation of the sparsity is achieved by the following sparsity ratio:
\begin{equation}
    \alpha =  1-\frac{\left \| x \right \|_0}{n \times (J\times Q)},
\end{equation}
where $\left \| . \right \|_0$ denotes the $l_0$ norm. It is clear that the closer the ratio is to $1$, the sparser the representation is. On the other hand, if the ratio equals zeros, none of the coefficients within the representation are null. 
\begin{figure}[H]
        \centering
            \begin{subfigure}[b]{0.475\textwidth}   
            \centering 
            \includegraphics[width=\textwidth]{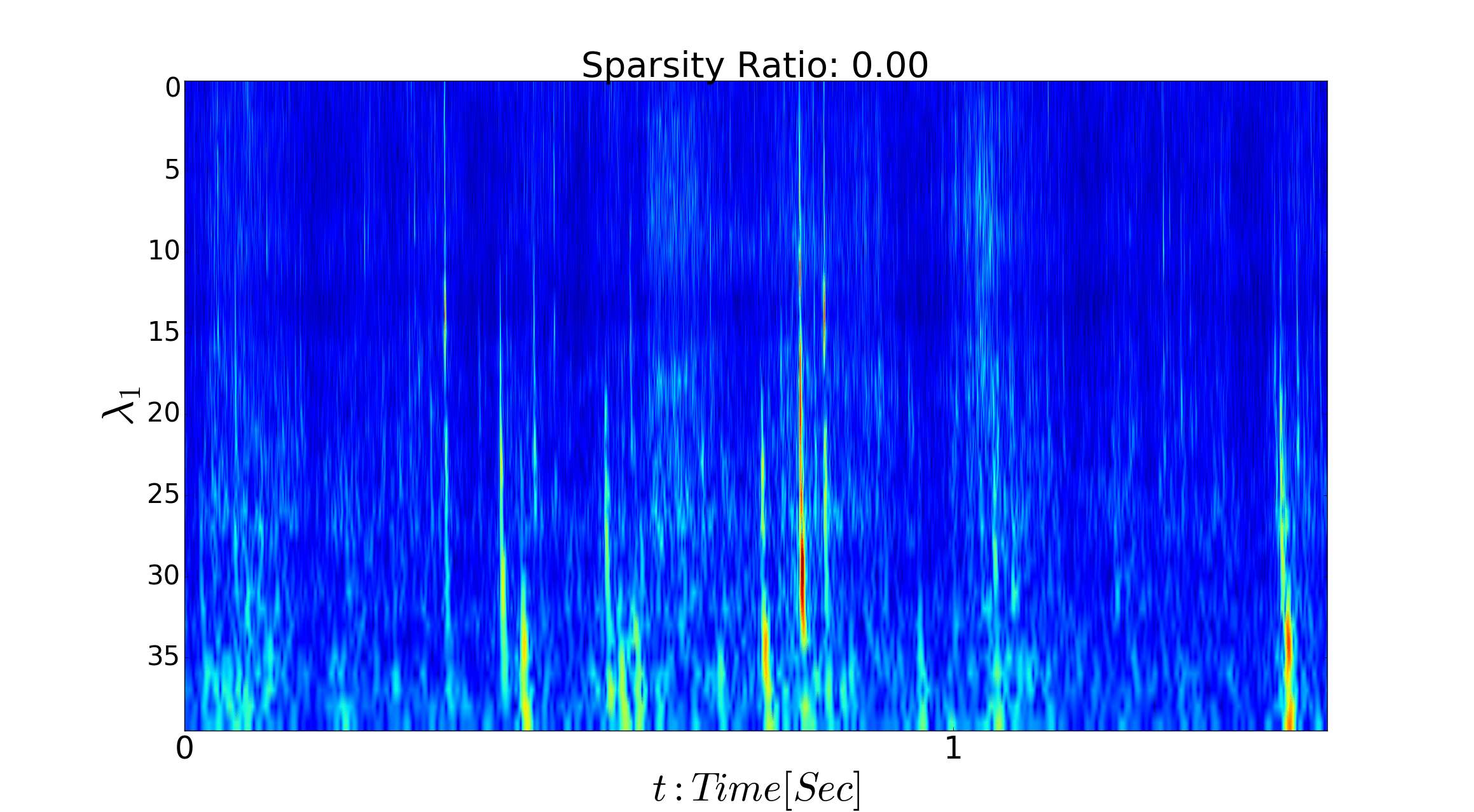}
            \caption{118596 - $\alpha=0.0 $}     
        \end{subfigure}
        \quad
        \begin{subfigure}[b]{0.475\textwidth}   
            \centering 
            \includegraphics[width=\textwidth]{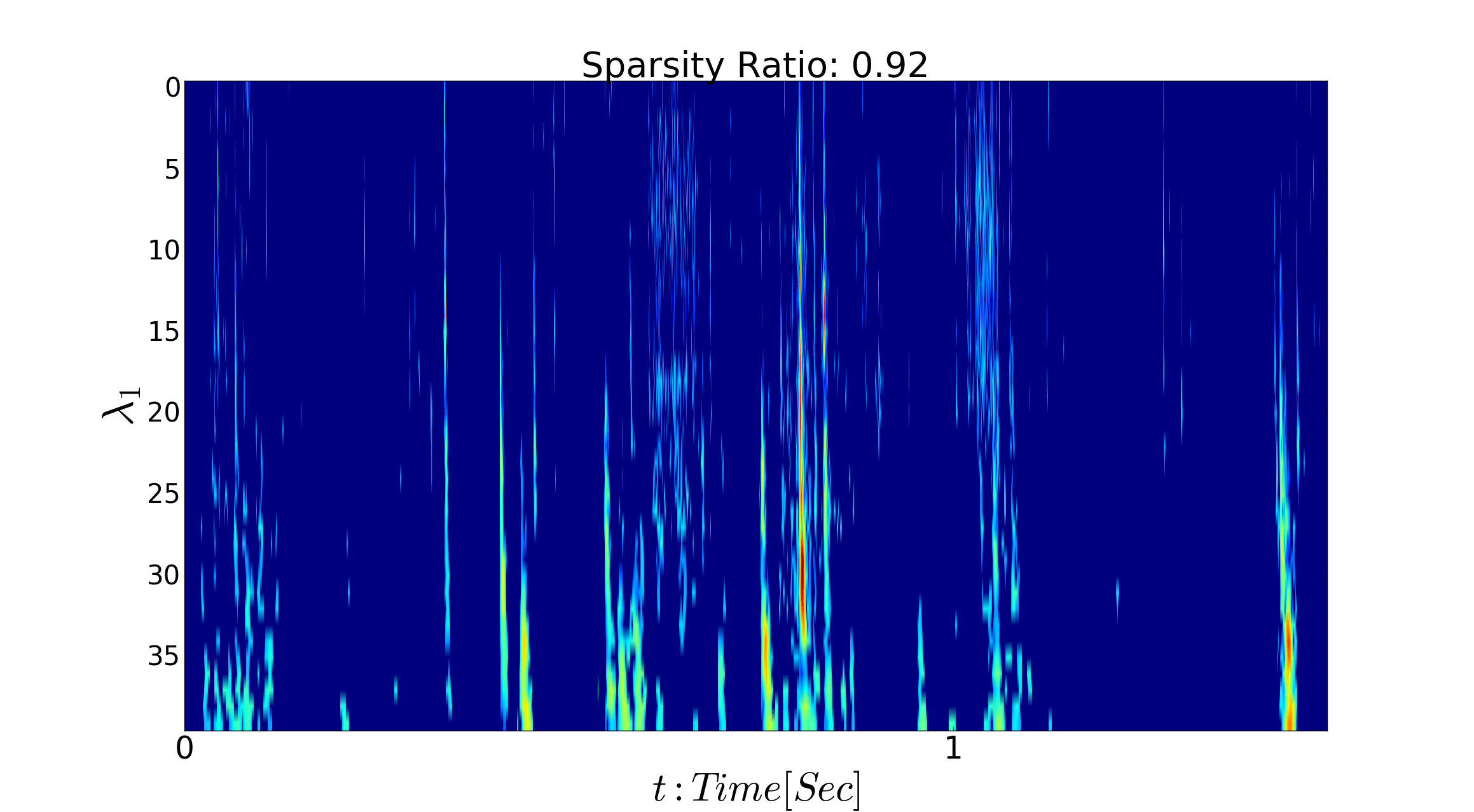}
            \caption{118596 Denoised - $\alpha=0.92 $}     
        \end{subfigure}
        \begin{subfigure}[b]{0.475\textwidth}   
            \centering 
            \includegraphics[width=\textwidth]{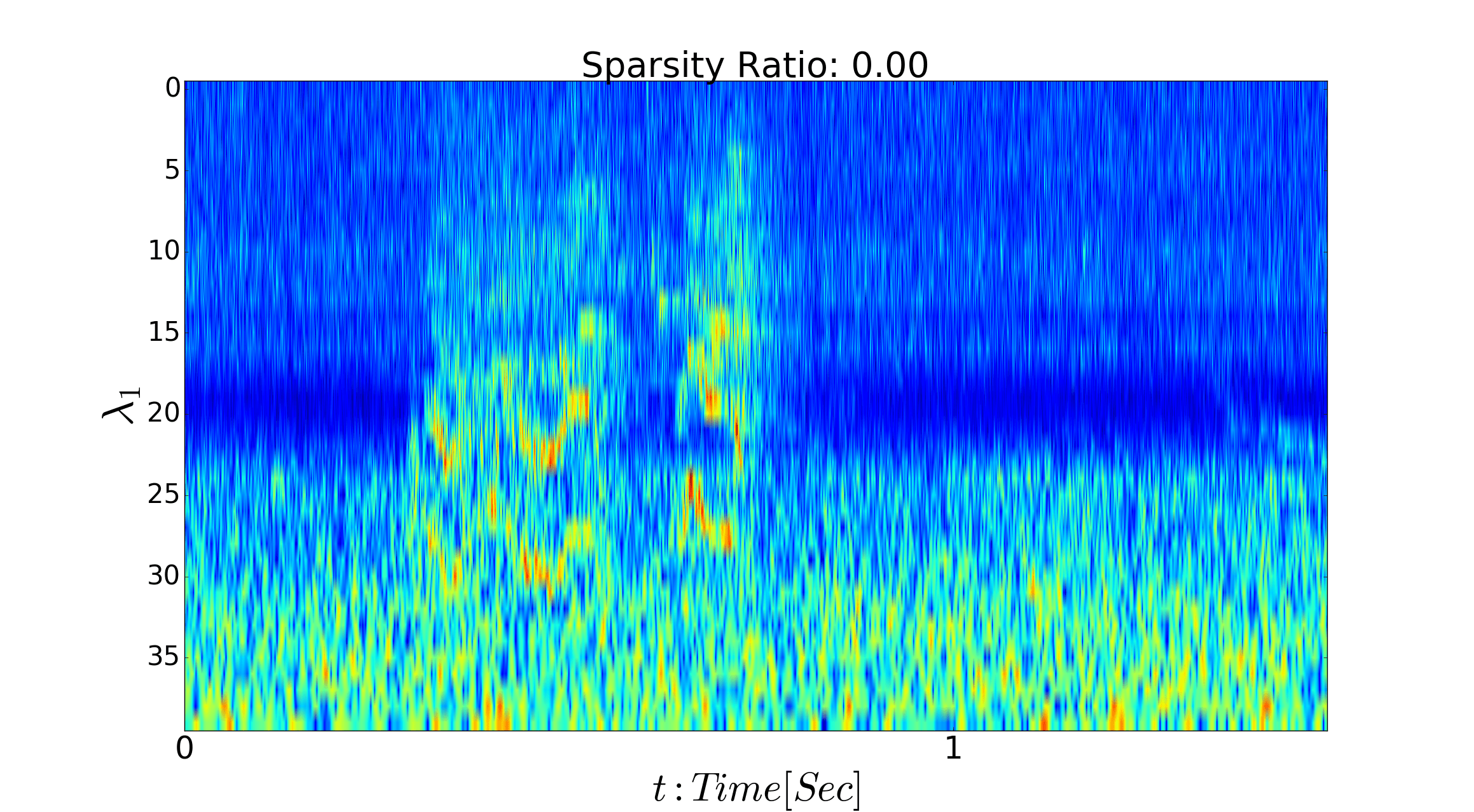}
            \caption{111092 - $\alpha=0.0$}     
        \end{subfigure}
        \quad
        \begin{subfigure}[b]{0.475\textwidth}   
            \centering 
            \includegraphics[width=\textwidth]{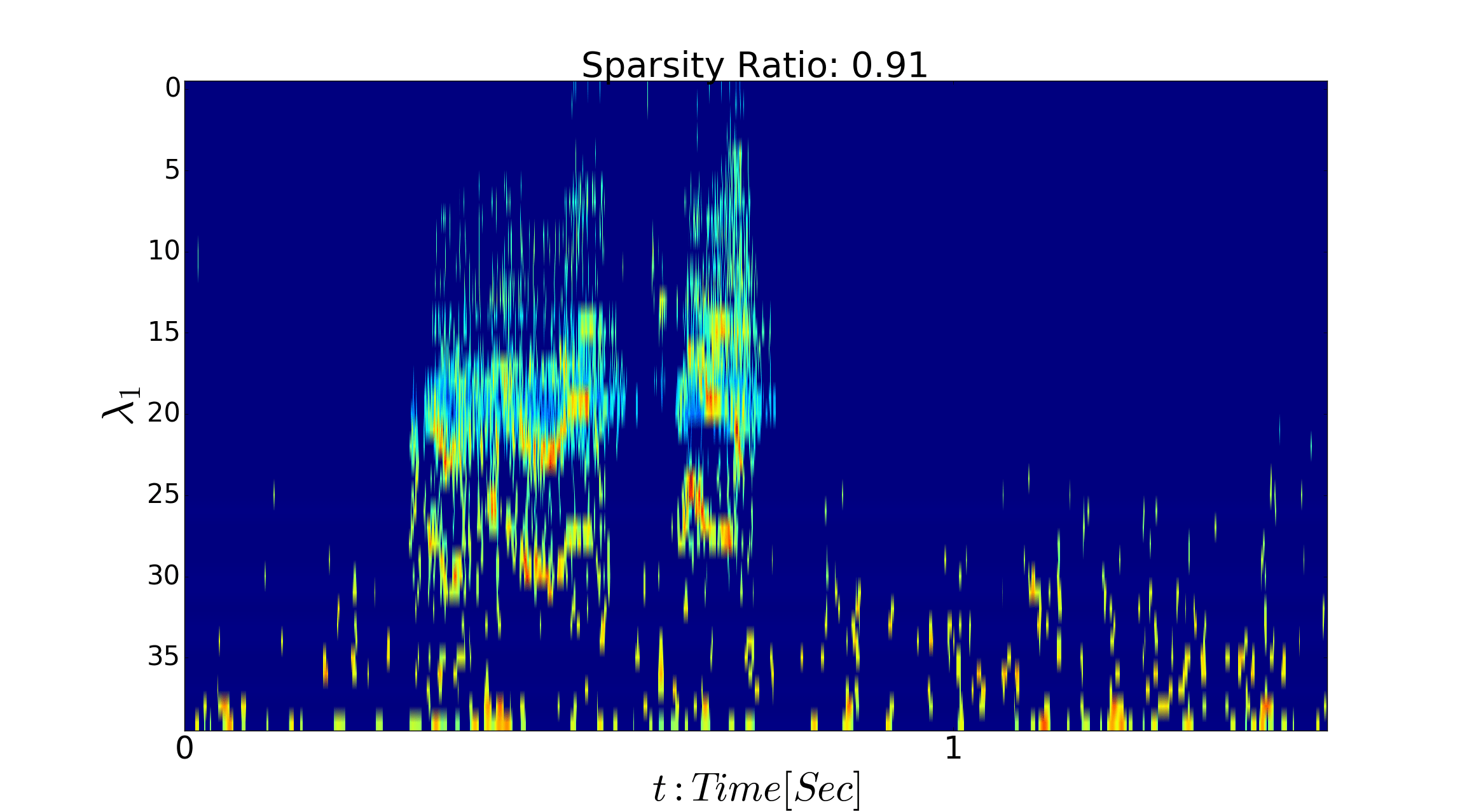}
            \caption{111092 Denoised - $\alpha=0.91 $}  
            \label{11092-d}
        \end{subfigure}
                \centering
        \begin{subfigure}[b]{0.475\textwidth}   
            \centering 
            \includegraphics[width=\textwidth]{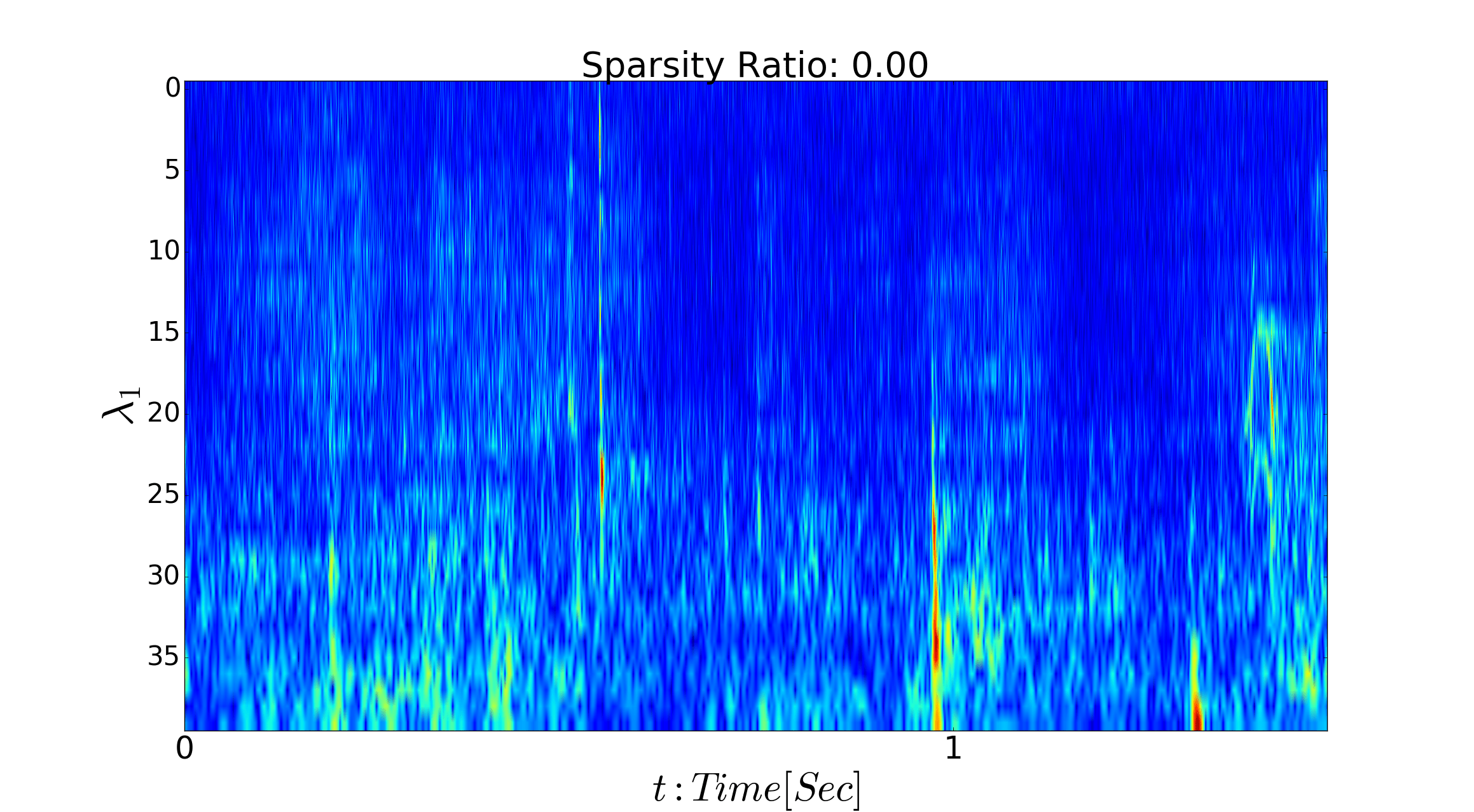}
            \caption{178878 - $\alpha=0.0 $}    
        \end{subfigure}
        \quad
        \begin{subfigure}[b]{0.475\textwidth}   
            \centering 
            \includegraphics[width=\textwidth]{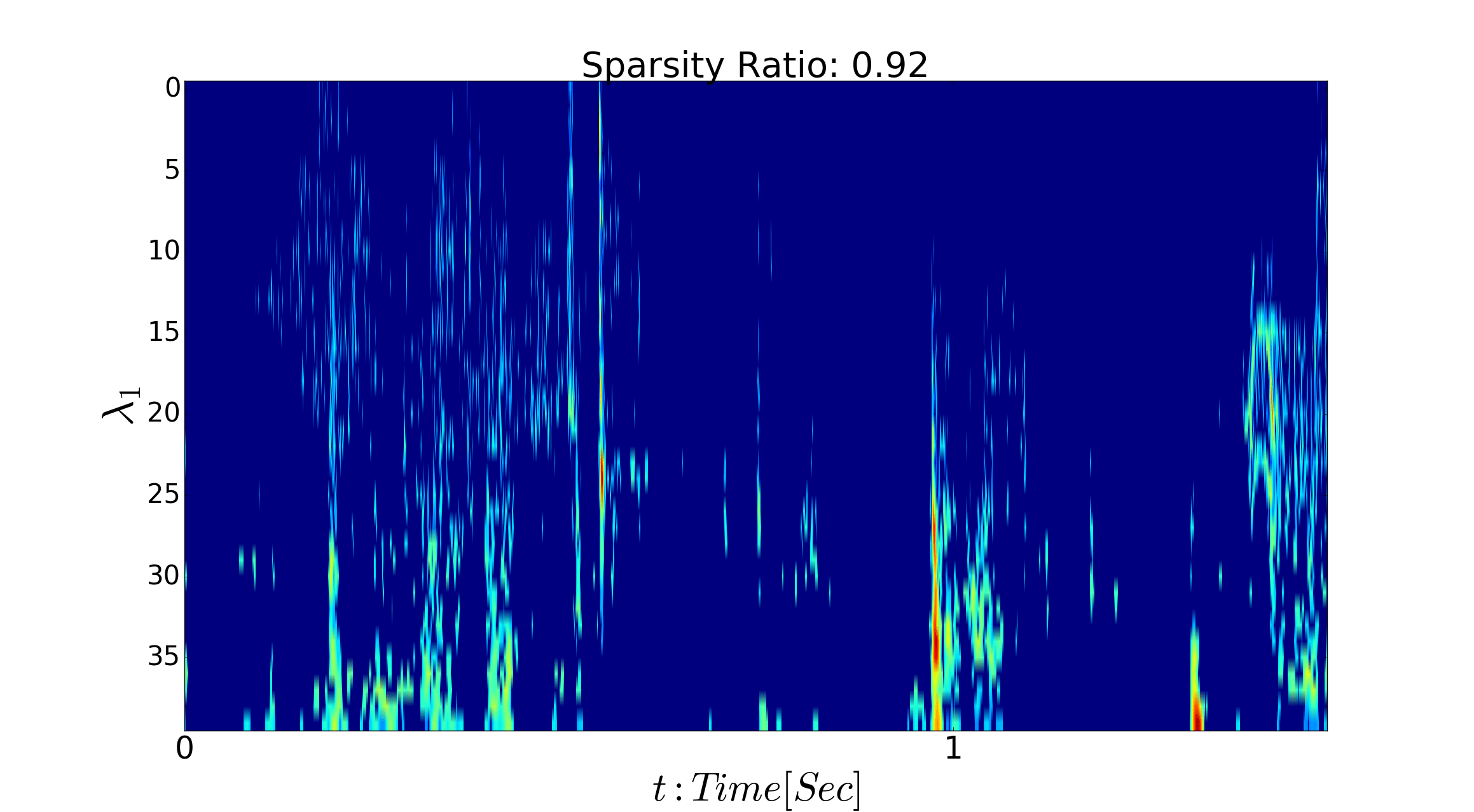}
            \caption{178878 Denoised - $\alpha=0.92 $}    
        \end{subfigure}
\end{figure}

        \begin{figure}[H]\ContinuedFloat

        \begin{subfigure}[b]{0.475\textwidth}   
            \centering 
            \includegraphics[width=\textwidth]{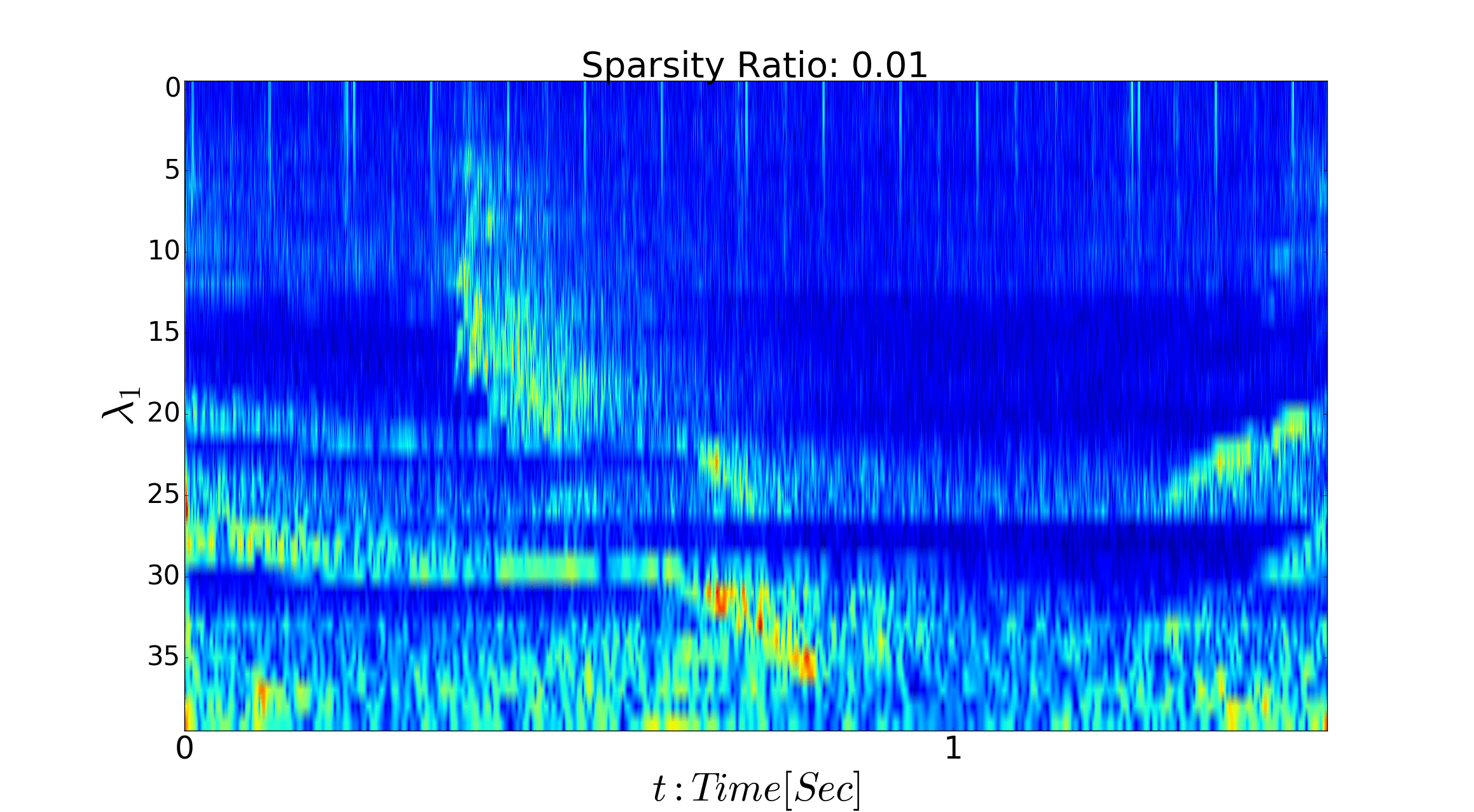}
            \caption{15657 - $\alpha=0.01 $}     
        \end{subfigure}
        \quad
        \begin{subfigure}[b]{0.475\textwidth}   
            \centering 
            \includegraphics[width=\textwidth]{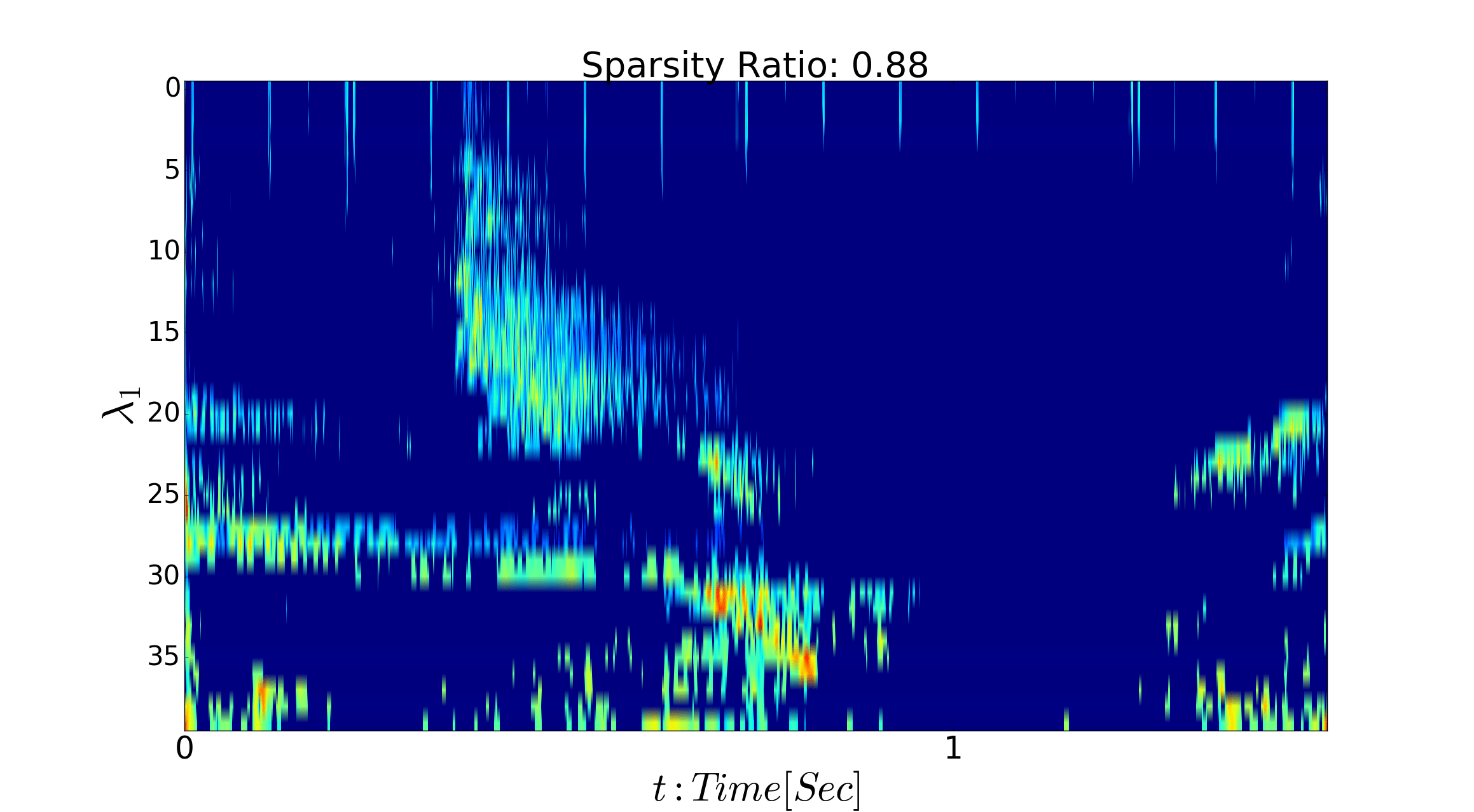}
            \caption{15657 Denoised - $\alpha=0.88 $}     
        \end{subfigure}
    \end{figure}
    \begin{figure}[H]\ContinuedFloat
        \centering
        \begin{subfigure}[b]{0.475\textwidth}   
            \centering 
            \includegraphics[width=\textwidth]{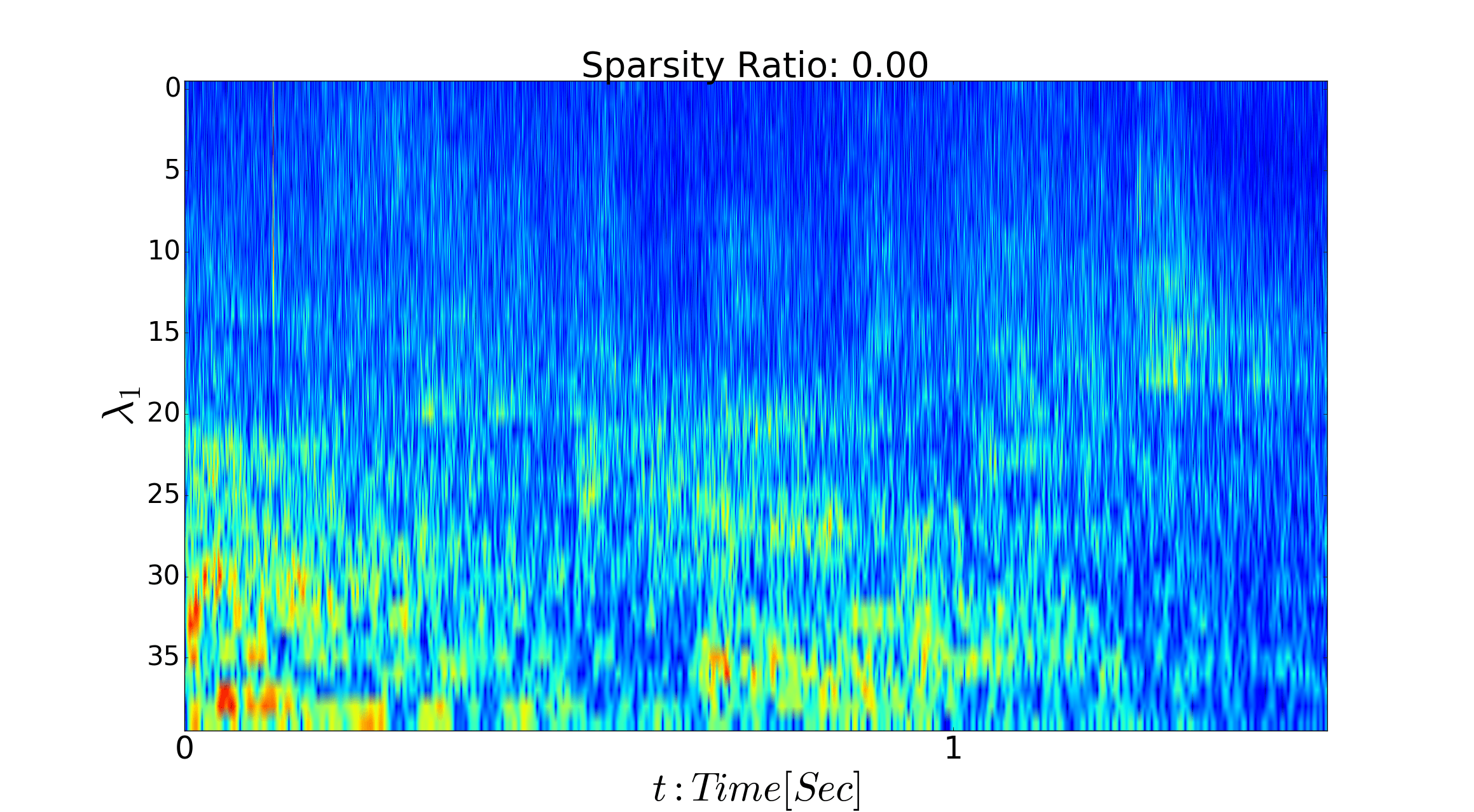}
            \caption{126154 - $\alpha=0.0 $}     
        \end{subfigure}
        \quad
        \begin{subfigure}[b]{0.475\textwidth}   
            \centering 
            \includegraphics[width=\textwidth]{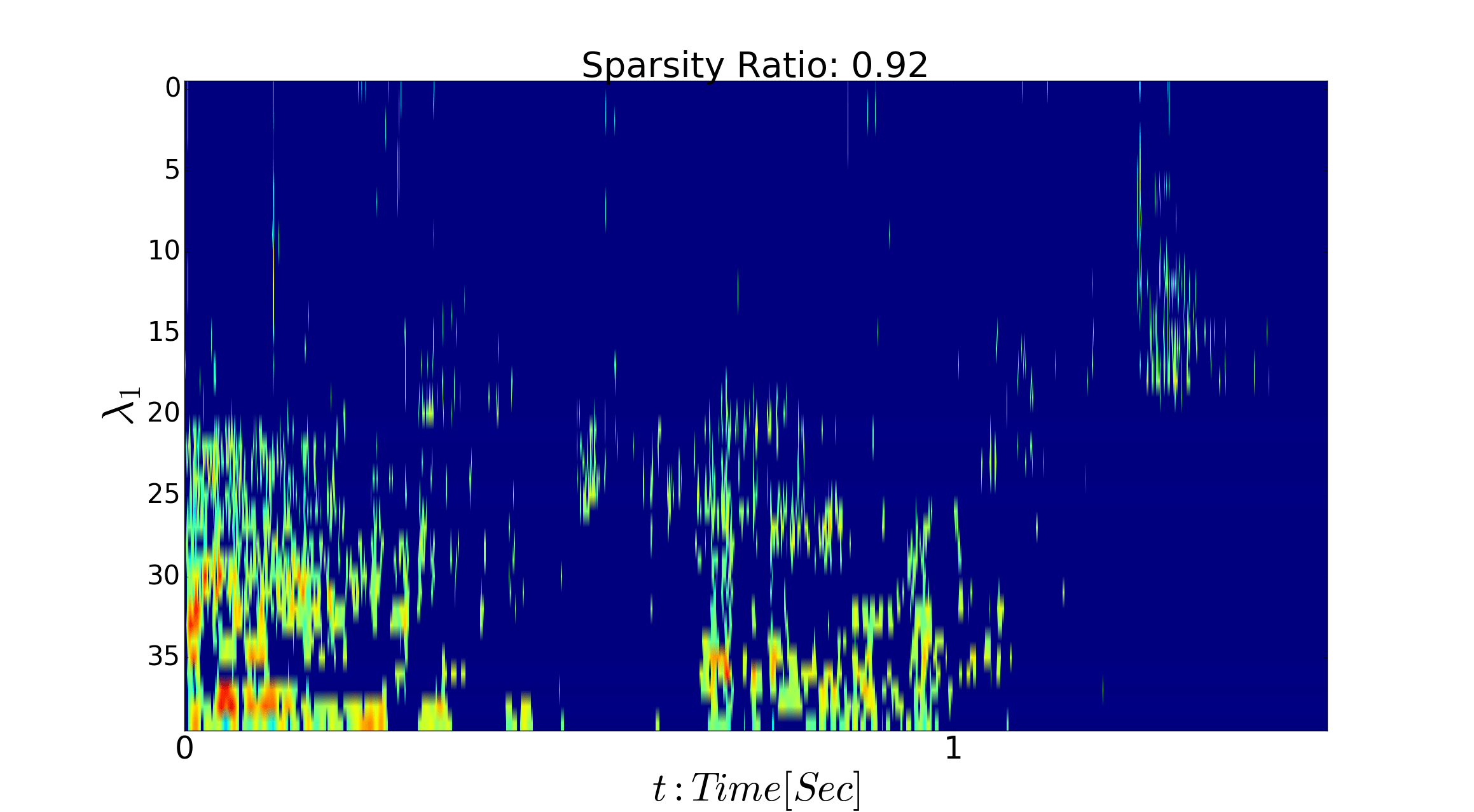}
            \caption{126154 Denoised - $\alpha=0.92 $}     
        \end{subfigure}
                \caption{ \textbf{Noisy Scalogram} ($U^1$) vs \textbf{Sparsity Evaluation of Denoised ($T[U^1]$)} - the wav name and the sparsity ratio are shown in each representation's title}
        \label{fig:sparse}
    \end{figure}

Given the fact that the data set contains recording via mobile application in diverses audio scenes, it is clear that many signals contain energy within all frequency bands. As such, the wavelet decomposition, even if non-orthogonal, is not sparse in most cases. Our denoising technique has the ability to remove the background noise and conserves only the foreground activity of each signal. In the special case of noise activity, one can see in Fig.\  \ref{11092-d}, that our algorithm conserves the bird song's chirp while removing almost all the other information.

\subsection{Quantitative Analysis: Sparse Deep Scattering Network}
\label{DeepCroisee}
We now propose a quantitative analysis via solving the supervised classification task by using a variant of the standard scattering network including our proposed denoising scheme.
The Deep Scattering Network, first developed in \cite{mallat2012group} and first successfully applied in \cite{bruna2011classification,anden2011multiscale} is based on a cascade of linear and non-linear operators on the input signal. The linear transformation is a wavelet transform, and the nonlinear transformation is a complex modulus. At each layer, the convolution of the representation with a scaling function leads to the so-called scattering coefficients.
This network is stable (Lipschitz-continuous) and suitable for machine learning tasks as it removes spatio-temporal nuisances by building space/time-invariant features.
The translation invariance property is provided by the scaling function that acts as an averaging operator on each layer of the transform leading to an exponential decay of the scattering coefficients \cite{waldspurger2017exponential}. Since the continuous wavelet transform increases the number dimension, the complex modulus is used as its contractive property reduces the variance of the projected space \cite{mallat2016understanding}.

The first layer of a DSN corresponds to standard scalogram, the second considers each frequency dimension of this scalogram as the input signal of the second layer. All the features are computed by applying a low-pass filter on each of these representations. In the SDSN, a non-linear thresholding operator is applied on each representation before to process both, the next layer as well as the scattering coefficient of the current layer.

\subsubsection{Sparse Deep Scattering Network}
\label{network}
\begin{figure}[h]
    \centering
    \includegraphics[width=\linewidth]{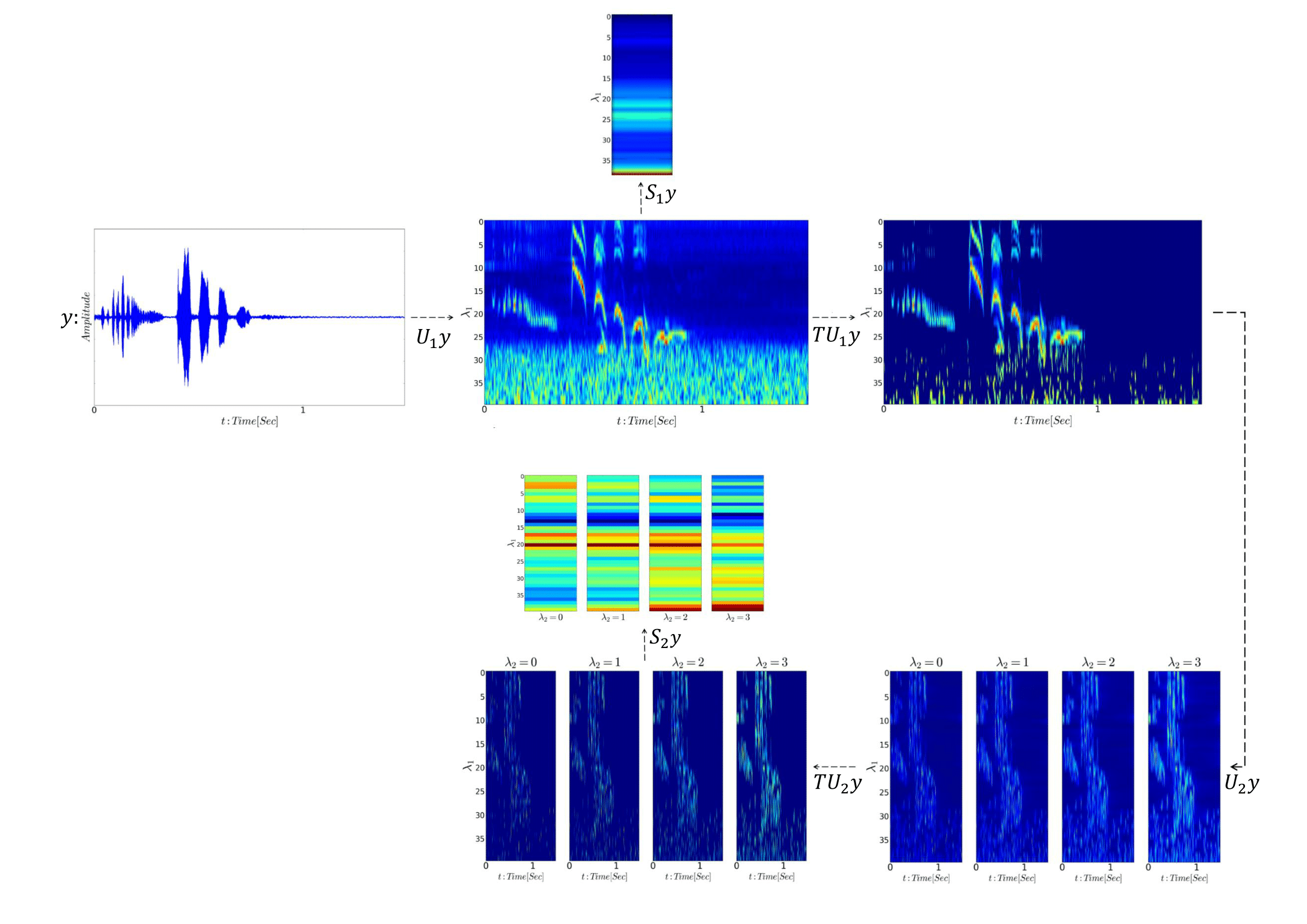}
    \caption{\textbf{Sparse Deep Scattering Network -Gammatone Wavelet} representation - J1= 5, Q1=8, J2=4, Q2=1}
    \label{fig:DS-apply}
\end{figure}

We denote by $U^{(1)}$ the application of the filter-bank of the first layer onto a signal $y$ corresponding to the first layer of the scattering network. The output of this first layer and as previously mentioned consists of a $2D$ tensor of shape $(J^{(1)}Q^{(1)},N)$ with $N$ the length of the input signal.
We omit here boundary conditions, sub-sampling, and consider a constant shape of $N$ throughout the representations.
We thus obtain,
\begin{align}
U^{(1)}[y]=|y \star \mathcal{W}^{(1)}|,
\end{align}
where $|.|$ operator corresponds to an element-wise complex modulus application.
As previously mentionned, we define the convolution operation between those two objects as,
\begin{align}
y \star \mathcal{W}^{(1)}= \begin{pmatrix}
y \star \mathcal{W}^{(1)}(1,.) \\
\dots \\
y \star \mathcal{W}^{(1)}(J^{(1)}Q^{(1)},.) 
\end{pmatrix}.
\end{align}
Then, we define by $\mathcal{T}$ the thresholding operator minimizing the empirical risk,
\begin{equation}
      \mathcal{T} = \arg \min_{\delta} \tilde{\mathcal{R}}(y,W^{(1)}).
\end{equation}
This non-linearity is applied to the latent representation $U^{1}$ producing the following sparse representation: $\mathcal{T}[U^{(1)}[y]]:=U^{(1)}_T[y]$.
From this sparse representation, the scattering coefficients can be computed leading to 
\begin{equation}
 S^{(1)}  = U^{(1)}_T[y] \star \phi^{(1)},
\end{equation}
$\phi^{(1)}$ being a ''low-pass'' filter. The application of the latter is performed for each row of the representation.
This application of a low frequency band-pass filter allows for symmetries invariances, inversely proportional to the cut-off frequency of $\psi$.
From this, the second layer is computed. 
We now denote the second layer representation as $U^{(2)}[U_T^{(1)}[y]]$  being
 a $3D$ tensor. 
 It defined as
\begin{align}
U^{(2)}[U_T^{(1)}[y]](j_2,.,.)= | U^{(1)}_T[y] \star \mathcal{W}^{(2)}(j_2)|.
\end{align}
Then, the thresholding operator is applied to each new representation $U^{(2)}[U_T^{(1)}[y]](.,j_1,.)$ independently. The final sparse representation is thus denoted by  $U_T^{(2)}[U_T^{(1)}[y]]$.
Given this $3D$ representation, the second layer scattering coefficients are defined as
\begin{equation}
S^{(2)}=U_T^{(2)}[U_T^{(1)}[y]] \star \phi^{(2)},
\end{equation}
We present an illustration of the network computation in Fig.\  \ref{fig:DS-apply}.

As can be seen in the proposed example, while the first layer provides time-frequency information, the second layer characterizes transients as demonstrated in \cite{balestriero2016robust}. With this extended framework, we now dive into the problem of thresholding redundant frame, cases where the quality factor, $Q$, is greater than $1$ which are in practice needed to bring enough frequency precision. 
We provide in Fig.\  \ref{fig:DS-apply} illustration showing the effect of each SDSN layers. It is clear that even if the first layer representation has been denoised, the second layer's latent representation, by decomposing each of the $U^{(1)}_T$ frequential dimension, still contains inherent noise. This noise is removed as it can be observed in the second thresholded layer representation $U^2_T$. 

As we will demonstrate below, our method as well as each contribution taken independently and jointly lead to significant increase in the final performance. We compare our results against the DSN. In all cases, the scattering coefficients are then fed into a random forest \cite{breiman2001random} with parameters\footnote{n\_estimator: $100$, min\_samples\_split:  $130$ ,class\_weights:'balanced\_subsample'}  based on the sklearn library \cite{pedregosa2011scikit}.

\begin{table*}[h!]
\centering
\caption{Classification Results - Bird Detection - Area Under Curve metric (AUC)}
\begin{tabular}{llll}
\hline
\multicolumn{1}{l|}{}              & $\textbf{\textit{min}}$& \begin{tabular}[c]{@{}l@{}}$\textbf{\textit{mean}}$\end{tabular}& \begin{tabular}[c]{@{}l@{}}$\textbf{\textit{max}}$\end{tabular} \\ \hline
\multicolumn{1}{l|}{\textbf{Sparse Deep Scattering Network}}   \\ \hline  
\multicolumn{1}{l|}{Gammatone}        & $\textbf{72.25}$     &$\textbf{74.23}$ & $\textbf{77.18}$\\
\multicolumn{1}{l|}{Morlet}        & $71.12$     &$73.49$ & $76.11$  \\
\hline
\multicolumn{1}{l|}{\textbf{Deep Scattering Network}}          \\ \hline
\multicolumn{1}{l|}{Gammatone}        & $70.69$     &$72.77$ & $75.82$\\
\multicolumn{1}{l|}{Morlet}        & $70.23$     &$72.37$ & $75.13$  \\
\hline
\end{tabular}
\label{table:compare}
\end{table*}

\subsubsection{Controlled Denoising Experiment}
We based our implementation on \cite{balestriero2017linear} leveraging the Fourier based computations of localized filter in the frequency domain. For all input signals we perform a renormalization to have unitary energy. 
We propose complementary experiments in this section to further demonstrate the need for thresholding even in the context of stable representations offered by the scattering network. To do so we simulate an environment for which we combine denoised and noised samples in both the training and test sets with different mixture proportions.
Thus we propose in table \ref{score} the described experiment achieved via the Gammatone wavelet only: for both the SDSN and DSN models. 
We evaluate the performance of the classifier for different mixture configuration as we can see in the table \ref{score}. The row corresponds to the percentage of noisy data in the training set, and the columns are the percentage of noisy data in the test set. It is clear that the percentage of denoised data in each set is one minus the percentage of noisy data. Then, in the table \ref{improve}, we show the difference between the mean accuracy of the SDSN-gammatone with respect to each mixture mean accuracy. 
\begin{table}[h]
\centering
\small
\caption{DSN evaluation with respect to mixture noise-denoised - AUC with mean and standard deviation over $50$ runs.}
\begin{tabular}{l|llllll}
 Train $\setminus$Test & $0\%$ & $20\%$ &$40\%$ &$60\%$ &$80\%$  &$100\%$\\[1pt]
  \hline
  $0\%$  & \textbf{74.2}$\pm .9$  &$69.6 \pm 1.1$ &$64.7 \pm 1.0$ &$60.0 \pm 1.0$ &$55.2 \pm 1.3$ &$50.3 \pm 1.1$ \\[1pt]
  $20\%$   &$73.9 \pm .8$ &\textbf{73.3 $\pm$ .9} &$72.8 \pm 1.0$ &$72.2 \pm .9$ &$71.5 \pm .9$ &$70.9 \pm 1.0$ \\[1pt]
  $40\%$   &$73.6 \pm .8$  &$73.3 \pm .9 $ &\textbf{73.1 $\pm$ .9} &$72.7 \pm 1.0$ &$72.3 \pm 1.0$ &$72.0 \pm .8$ \\[1pt]
  $60\%$ &$73.0 \pm .9$    &$73.1 \pm 1.$ &$72.8 \pm .9$ &\textbf{72.7 $\pm$ .9} &$72.7 \pm .8$ &$72.4 \pm .9$ \\[1pt]
  $80\%$ &$71.8 \pm 1.$    &$71.8 \pm 1.1$ &$72.3 \pm 1.0$ &$72.4 \pm 1.0$ &\textbf{72.5 $\pm$ .9} &$72.8 \pm .9$ \\[1pt]
  $100\%$  &$50.0 \pm .0$  &$54.5 \pm .6$ &$59.2 \pm .8$ &$63.7 \pm .9$ &$68.3 \pm .9$ &\textbf{72.77 $\pm$ .8 } \\[1pt]
\end{tabular}
\label{score}
\end{table}
\begin{table}[h]
\centering
\caption{SDSN - Improve gain in term of mean values - AUC}
\begin{tabular}{l|llllll}
Train $\setminus$ Test & $0\%$ & $20\%$ &$40\%$ &$60\%$ &$80\%$  &$100\%$\\[1pt]
  \hline
  $0\%$ & $+0$  &$+4.6$ &$+9.5$ &$+14.2$ &$+19.0$ &$+23.9$ \\[1pt]
  $20\%$ & $+.38$  &\textbf{+0.9} &$+1.43$ &$+2.0$ &$+2.73$ &$+3.3$ \\[1pt]
  $40\%$ &  $+.68$   &$+.9$ &\textbf{+1.13} &$+1.53$ &$+1.93$ &$+2.23$ \\[1pt]
  $60\%$ &$+1.28$    &$+1.13$ &$+1.43$ &\textbf{+1.53} &$+1.53$ &$+1.82$ \\[1pt]
  $80\%$ &$+2.48$    &$+2.43$ &$+1.93$ &$+1.0$ &\textbf{+1.73} &$+1.43$ \\[1pt]
  $100\%$ &  $+24.28$ &$+19.7$ &$+15.0$ &$+10.5$ &$+5.9$ &\textbf{+1.5} \\[1pt]
\end{tabular}
\label{improve}
\end{table}
The configuration depicted in the diagonal of the tables \ref{score} and \ref{improve} corresponds to the ''ideal'' case of identical proportion of noise in the train and test set. However for many practical applications more volatile regimes can be witnessed. Those correspond either to the upper diagonal part for which the proportion of noisy samples is more important in the test set than the train set. Conversely the lower diagonal corresponds to the symmetric case. We can see that even in the ideal setting, introduction of the SDSN brings reasonable robustness to the classifier pipeline. More importantly for the off diagonal cases, a significant gain is achieved via the SDSN.



\bibliography{iclr2018_conference.bib}
\bibliographystyle{plain}
\newpage
\appendix
\section{Building a Deep Scattering Network}
\label{appendixA}
\subsection{Continous Wavelet Transform}\label{def}
\begin{center}
"By oscillating it resembles a wave, but by being localized it is a wavelet".\\ \raggedleft {Yves Meyer} 
\end{center}
Wavelets were first introduced for high resolution seismology \cite{cycle} \cite{grossmann1984decomposition} and then developed theoretically by Meyer et al. \cite{painless}.
Formally, wavelet is a function $\psi \in \mathbb{L}^2$ such that:
\begin{equation}
\int \psi(t) dt = 0,
\end{equation}
it is normalized such that $\left \| \psi \right \|_{\mathbb{L}^2} = 1$. 
Often, wavelets can be categorized in two distinct groups, the discrete wavelets and the continuous ones. The discrete wavelets are constructed based on a system of linear equation that represent the atom's property. These wavelet, when scaled in a dyadic fashion form an orthonormal basis.
On the other hand, the continuous wavelets have an explicit formulation and build an over-complete dictionary when successively scaled. In this work, we focus on the continuous wavelets as they provide a more complete tool for analysis of signals. 
In order to perform a time-frequency transform of a signal, we first build a filter bank based on the mother wavelet. This wavelet is names the mother wavelet since it will be dilated and translated in order to create the filters that will constitute the filter bank. Notice that wavelets have a constant-Q property, thereby the ratio bandwidth to center frequency of the children wavelets are identical to the one of the mother. Then, the more the wavelet atom is high frequency the more it will be localized in time.
The usual dilation parameters follows a geometric progression and belongs to the following set:
\[ \Lambda = \left \{ 2^{(j-1)/Q}, j= 1,...,J \times Q  \right \}.  \] 
Where the integers $J$ and $Q$ denote respectively the number of octaves, and the number of wavelets per octave.
Having selected a geometric progression ensemble, the dilated version of the mother wavelet in the time are computed as follows:   
    \[ \psi_{\lambda}(t)=\frac{1}{\lambda} \psi ( \frac{t}{\lambda} ),   \; \forall \lambda \in \Lambda,       \]
and can be calculated in the Fourier domain as follows:
\[  \hat{\psi}_{\lambda}(\omega) = \hat{\psi}(\lambda \omega), \; \forall \lambda \in \Lambda.            \]
Notice that in practice the wavelets are computed in the Fourier domain as the wavelet transform will be based on a convolution operation which can be achieved with more efficiency in the frequency plane. 
By construction, the children wavelets have the same properties than the mother one. As a result, in the Fourier domain:
\[ \hat{\psi}_{\lambda} = 0, \; \forall \lambda \in \Lambda. \]
Thus, to create a filter bank that cover all the frequency support, one needs a function that captures the low frequencies contents. The function is called the scaling function and satisfies the following criteria:
\[ \int \phi(t) dt = 1. \]


\subsection{Wavelet Families}
\label{family}
Among the continuous wavelets, different families exist. Each posses different properties, such as bandwidth, center frequency. This section is dedicated to the development of the families that are important for the analysis of diverse signals.
\subsubsection{The Morlet wavelet}
\begin{figure}[H]
    \centering
    \includegraphics[width=.9\linewidth]{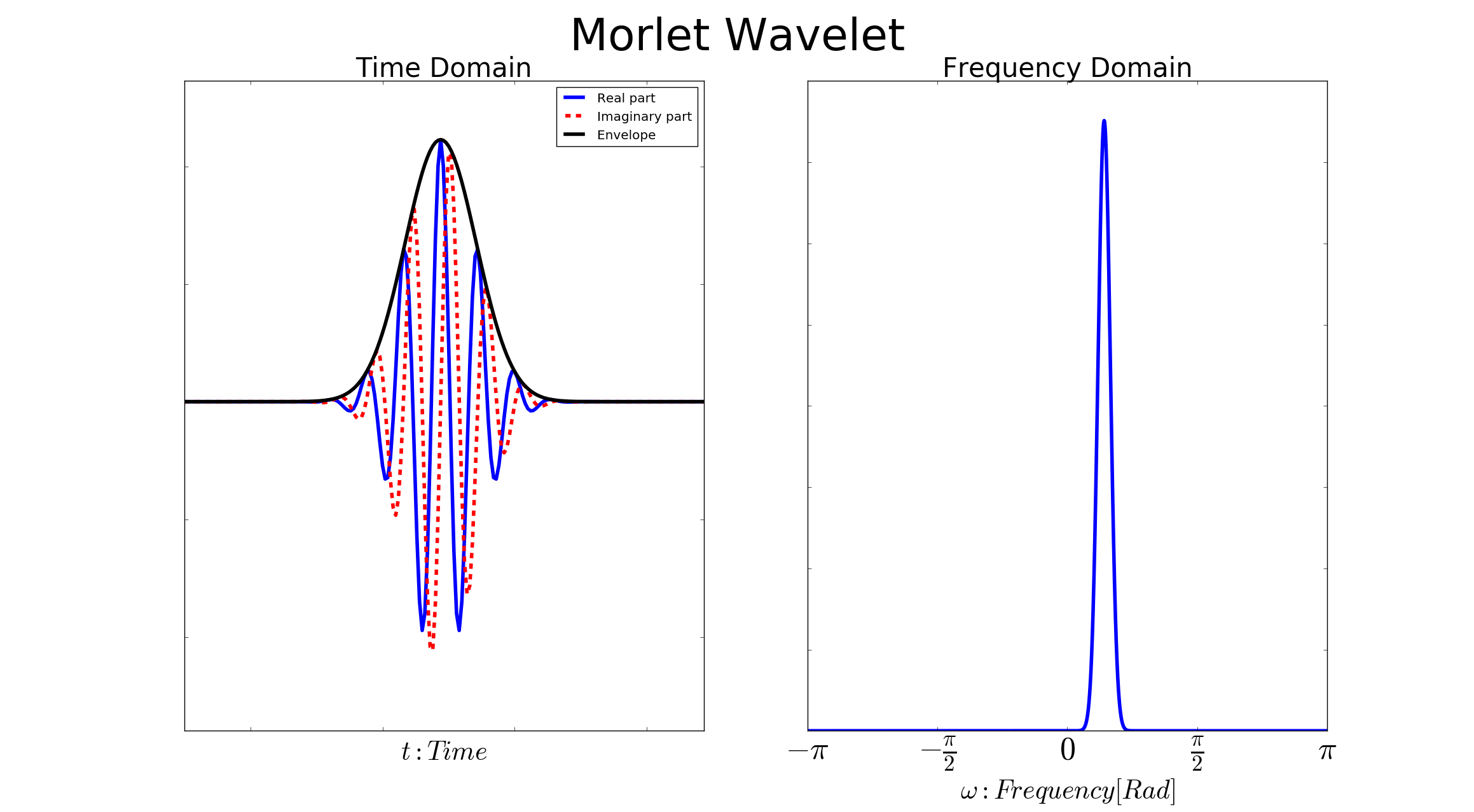}
    \caption{On the left a Morlet wavelet in the time domain where the dashed line is the imaginary part, the solid line is the real part, and the black envelope is the complex modulus, on the right a Morlet wavelet in the frequency domain.}
    \label{fig:Morlet}
\end{figure}

The Morlet wavelet (Fig.\  \ref{fig:Morlet}) is built by modulating a complex exponential and a Gaussian window defined in the time domain by,
\begin{equation}
\label{Morlet}
\psi^{\textit{M}}(t) = \pi^{-\frac{1}{4}} e^{i\omega_0 t}e^{-\frac{t^2}{2}},
\end{equation}
where $\omega_0$ defines the frequency plane.
In the frequency domain, we denote it by $\hat{\psi}^{\textit{M}}(t)$,
\begin{equation}
\hat{\psi}^{\textit{M}}(\omega) = \pi^{-\frac{1}{4}}e^{-\frac{(\omega-\omega_0)^2}{2}}, \forall \omega \in \mathbb{R}^{\star}_{+},
\end{equation}
thus, it is clear that $\omega_0$ defines the center frequency of the mother wavelet.
With associated frequency center and standard deviation denoted respectively by $\omega_{c}^{\lambda_i}$ and $\Delta^{\lambda_i}\omega$, $\forall j \in \{0,...,J*Q-1 \}$ are:
\begin{align*}
    \omega_{c}^{\lambda_i}&= \frac{\omega_0}{\lambda_i} \nonumber, \\ 
    \Delta^{\lambda_i}\omega&= \frac{1}{2 \lambda_i^2} \nonumber.
\end{align*}
Notice that for the admissibility criteria $\omega_0 = 6$, however one can impose that zeros-mean condition facilely in the Fourier domain. Usually, this parameter is assign to the control of the center frequency of the mother wavelet. Then, we are able to vary the parameter $\omega_0$ in order to have different support of Morlet wavelet. 
The Morlet wavelet, is optimal from the uncertainty principle point of view \cite{Wavtour}. The uncertainty principle, when given a time-frequency atoms, is the area of the rectangle of its joint time-frequency resolution. In the case of wavelet, given the fact that their ratio bandwidth to center frequency are equal, it implies that this area is equal for the mother wavelets and its scaled versions. As a result, because of its time-frequency versatility this wavelet is wildly used for biological signals such as bio-acoustic \cite{balestriero2015scattering}, seismic traces \cite{chopra2015choice}, EEG \cite{dwavelet} data.

\subsubsection{The Gammatone wavelet}
\begin{figure}[H]
\centering
\begin{subfigure}[b]{.8\textwidth}
    \includegraphics[width=.9\textwidth]{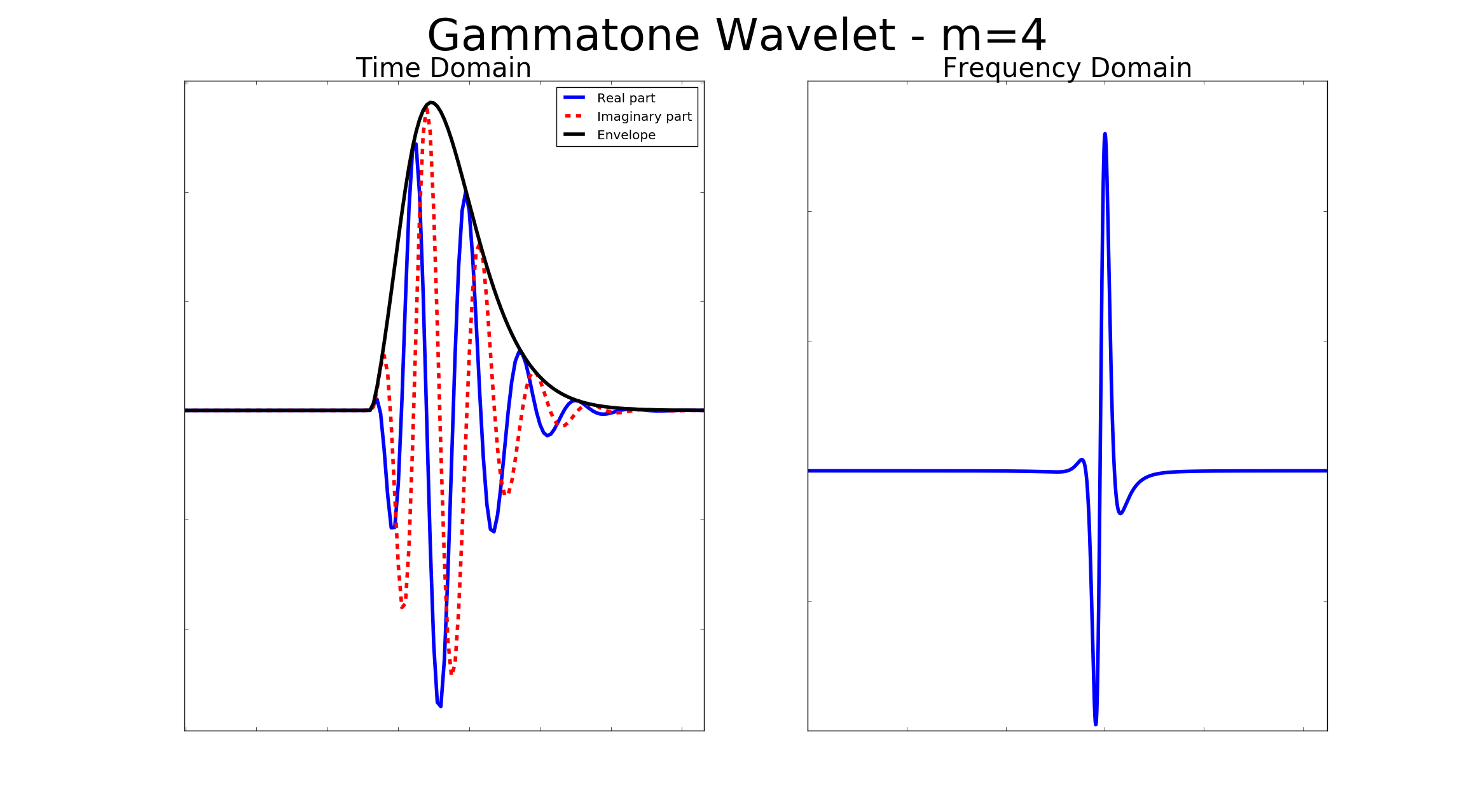}
\end{subfigure}
~
\begin{subfigure}[b]{.8\textwidth}
    \includegraphics[width=.9\textwidth]{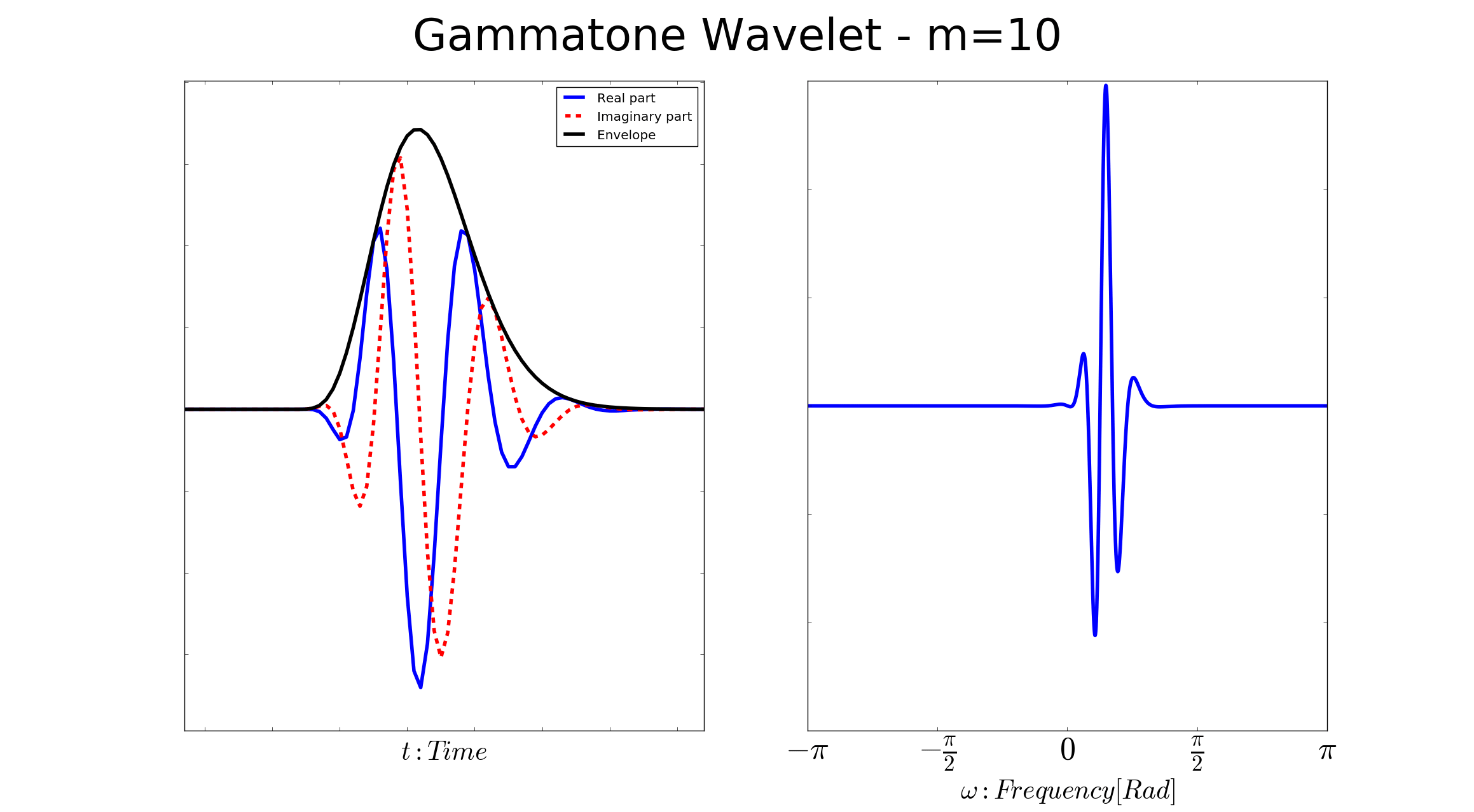}
\end{subfigure}
\caption{On the upper (bottom) left a $m=4$ ($m=10$) Gammatone wavelet in the time domain where the dashed line is the imaginary part and the solid line is the real part, on the upper (bottom) right a $m=4$  ($m=10$) wavelet in the frequency domain.}\label{fig:Gammatone}
\end{figure}

The Gammatone wavelet is a complex-valued wavelet that has been developed by \cite{venkitaraman2014auditory} via a transformation of the real-valued Gammatone auditory filter which provides a good approximation of the basilar membrane filter \cite{flanagan1960models}. Because of its origin and properties, this wavelet has been successfully applied for classification of acoustic scene \cite{lostanlenbinaural}.
The Gammatone wavelet (Fig.\  \ref{fig:Gammatone}) is defined in the time domain by,
\begin{equation}
\label{Gammatone}
\psi^{\textit{G}}(t) = \left ( 2 \pi(i-\sigma)t^{m-1}+(m-1)t^{m-2}\right )e^{-2pi\sigma t}e^{2 pi i t},
\end{equation}
and in the frequency domain by,
\begin{equation}
\hat{\psi}^{\textit{G}}(\omega) =  \frac{i \omega (m-1)!}{\left (\sigma +i(\omega-\sigma) \right )^m}.
\end{equation}
A precise work on this wavelet achieved by V. Lostalnen in \cite{lostanlen2017operateurs} allows us to have an explicit formulation of the parameter $\sigma$ such that the wavelet can be scaled while respecting the admissibility criteria:
\begin{align*}
\sigma^2 = \frac{ r^{\frac{2}{m}} (1 - r^{\frac{2}{m}}) m^2 \xi^2}{2} \left ( \sqrt{1+ \frac{B^2}{(1 - r^{\frac{2}{m}})^2 m^2 \xi^2}} -1 \right ),
\end{align*}
where $\xi$ is the center frequency and $B$ is the bandwidth parameter. Notice that $B=(1-2^{-\frac{1}{Q}}) \xi$ with $\xi= \frac{2 \pi}{1+ 2^{\frac{1}{Q}}}$ induce a quasi orthogonal filter bank.
The associated frequency center and standard deviation denoted respectively by $\omega_{c}^{\lambda_i}$ and $\Delta^{\lambda_i}\omega$, $\forall j \in \{0,...,J*Q-1 \}$ are thus:
\begin{align*}
    \omega_{c}^{\lambda_i}&= \xi \nonumber, \\ 
    \Delta^{\lambda_i}\omega&= B \nonumber.
\end{align*}
For this wavelet, thanks to the derivation in \cite{lostanlen2017operateurs}, we can manually select for each order $m$ the center frequency and bandwidth of the mother wavelet, which ease the filter bank design.
An important property that is directly related to the auditory response system is the asymmetric envelope, thereby the Gammatone wavelet is not invariant to time reversal to the contrary of the Morlet wavelet that behaves as a Gaussian function. Thus, for tasks such as sound classifications, this wavelet provides an efficient filter that will be prone to perceive the sound's attacks. Beside this suitable property for specific analysis, this wavelet is near optimal with respect to the uncertainty principle. Notice that, when $m \rightarrow \infty$ it yields the Gabor wavelet \cite{Cohen:1995}. Another interesting property of this wavelet is the causality, by taking into account only the previous and present information, there is no bias implied by some future information and thus it is suitable for real-time signal analysis.

\section{Thresholding}
\label{appendixC}
\subsection{Thresholding in an orthogonal basis}
Assuming that the observed signal $y$, is corrupted with white noise, 
 \begin{equation}
 y = x+ \epsilon,
 \end{equation}
 where $\epsilon$ is a vector of i.i.d centered normal distributions $\mathcal{N}(0,\sigma^2)$. 
 Now we define the estimate of $x$ by $\hat{x}_{W,D}$ such that:
 \begin{equation}
     \hat{x}_{W,D}(y) = W^{T}D^{S}Wy
 \end{equation}
 where $W$ denotes the orthogonal basis and $D^{S}$ is a diagonal binary operator such that,
 \begin{equation}
 D^{S}_{i,i}= \delta_i = \left\{\begin{matrix}
 1  \: \: if \: \: i \in S  \\ 
 0  \: \: if \: \: i \in U 
\end{matrix}\right. 
\end{equation},
 where $U$ and $S$ denote respectively the set of selected and unselected wavelet coefficients. We also define $D^{U}$ such that $I = D^{U} + D^{S}$.
 This estimate corresponds to a thresholding operation in the new basis and the inverse transform of this truncated representation. 

 We define the denoising problem as the solution of the following mean-square error:
 \begin{align}
 \mathcal{R}_{o}^{\star}(x,W)&=  \min_{\delta} \mathbb{E} \left \| x-\hat{x}_{W,D}(y) \right \|^2 \\
 & =   \min_{\delta} \mathbb{E} \left \| W^{T} (Wx-D^{S}Wy \right \|^2 \\
 & =  \min_{\delta} \mathbb{E} \left \|D^{U} Wx - D^{S}W(x+\epsilon) \right \|^2 \\
 & =  \min_{\delta} \left \| D^{U} Wx  \right \|^2 + \sigma^{2} tr (D^{S} WW^{T} D^{S}) \\
 & =  \min_{\delta} \sum_{i}^{n} [Wx]_i^{2} 1_{\left \{ \delta_i = 0 \right \}} + \sigma^{2} 1_{\left \{  \delta_i =1\right \}} \\
 & =  \sum_i^{n} \min ([Wx]_{i}^{2},\sigma^2).
 \end{align}
 Therefore, the optimal $D^{S^{\star}}$ and $D^{U^{\star}}$ given by the following $\delta$ values:
\begin{equation}
   \delta_i =  1_{\left | [Wx]^2_i \right |>\sigma^2}.
\end{equation}
\subsection{Upper-bound Non-orthogonal Risk \& Empirical Risk}
\label{upper-bound}

\begin{equation} 
     \hat{x}_{W,D}(y) = W^{\dagger}D^{S}Wy
 \end{equation}
\begin{align}
 \mathcal{R}^{\star}(x,W)&=  \min_{\delta} \mathbb{E} \left \| x-\hat{x}_{W,D}(y) \right \|^2 \\
 & =   \min_{\delta} \mathbb{E} \left \| W^{\dagger} (Wx-D^{S}Wy \right \|^2 \\
 & =  \min_{\delta} \mathbb{E} \left \|W^{\dagger} (D^{U} Wx - D^{S}W\epsilon) \right \|^2 \\
 & =  \min_{\delta} \left \| W^{\star}D^{U} Wx  \right \|^2 + \sigma^{2} tr (W^{T}D^{S} W^{\dagger}W^{\dagger^{T}} D^{S}W),
 \end{align}
Developing the previous expression and denoting by $\mu =  Wx$ the wavelet coefficient vector, we have:
\begin{align}
 \mathcal{R}^{\star}(x,W) = & \min_{\delta}  \sum_{i,j=1}^{n \times (J \times Q)} \mu_{i}\mu_{j} \psi^{\dagger^{T}}_{i} \psi^{\dagger}_{j} 1_{\left \{ \delta_i = 0, \delta_j = 0 \right \} } \nonumber \\
 & + \sigma^{2} \sum_{i,j=1}^{n \times (J \times Q)}  ( \psi^{\dagger^{T}}_{i} \psi^{\dagger}_{j} ) \psi_i^{T} \psi_j 1_{\left \{ \delta_i =1,\delta_j =1 \right \}}.
 \end{align}
we first use the triangular inequality,

\begin{align}
    \mathcal{R}^{\star}(x,W)  & \leq  \min_{\delta} \sum_{i,j=1}^{n \times (J \times Q)}  \left |     \mu_{i}\mu_{j} \psi^{\dagger^{T}}_{i} \psi^{\dagger}_{j}   \right | 1_{\left \{ \delta_i = 0, \delta_j = 0 \right \} } \nonumber \\
    & +  \sigma^{2} \sum_{i,j=1}^{n \times (J \times Q)}   \left | (\psi^{\dagger^{T}}_{i} \psi^{\dagger}_{j}) \psi_i^{T} \psi_j  \right | 1_{\left \{ \delta_i =1,\delta_j =1 \right \}}
\end{align}
Now let's,
\begin{equation}
\mathcal{R}^{U}[i,j]=   \left |    \mu_{i}\mu_{j} ( \psi^{\dagger^{T}}_{i} \psi^{\dagger}_{j})   \right |,
\end{equation}
and,
\begin{equation}
\mathcal{R}^{S}[i,j]= \sigma^{2}    \left |(\psi^{\dagger^{T}}_{i} \psi^{\dagger}_{j}) \psi_i^{T} \psi_j  \right |.
\end{equation}
Then, based on the following min-max formulation, we obtain an upper bound of the ideal risk, that, when minimized will approximate the ideal risk in the overcomplete case:
\begin{align}
     \mathcal{R}^{\star}(x,W)  & \leq  \sum_{k =1}^{n \times (J \times Q)} \min_{\delta_k} \sum_{j=1}^{n \times (J \times Q)} \max_{\delta_j, j\neq k} (1_{\left \{ \delta_k = 0, \delta_j = 0 \right \} } \mathcal{R}^{U}[k,j] + 1_{\left \{ \delta_k =1,\delta_j =1 \right \}}\mathcal{R}^{S}[k,j] ) \\
     & \leq  \sum_{k =1}^{n \times (J \times Q)} \min_{\delta_k}  \sum_{j=1}^{n \times (J \times Q)} (\max_{\delta_j, j\neq k} 1_{\left \{ \delta_k = 0, \delta_j = 0 \right \} } \mathcal{R}^U[k,j]  +  \max_{\delta_j, j\neq k}  1_{\left \{ \delta_k =1,\delta_j =1 \right \}}\mathcal{R}^S[k,j] ) \\
     & =  \sum_{k =1}^{n \times (J \times Q)} \min_{\delta_k}  \: \: 1_{\left \{ \delta_k =0 \right \}} \left [ \sum_{j =1}^{n \times (J \times Q)} \left |   \mu_{k}\mu_{j} \psi^{\dagger^{T}}_{k} \psi^{\dagger}_{j}   \right |\right ] \nonumber \\
     & \: + 1_{\left \{ \delta_k =1 \right \}}  \left [ \sigma^{2} \sum_{j=1}^{n \times (J \times Q)}   \left | (\psi^{\dagger^T}_{k} \psi^{\dagger}_{j}) \psi_k^{T} \psi_j  \right | \right ].
\end{align}
Now, let's denote by $\mathcal{R}_{up}^{U}$ the error term corresponding to unselected coefficients:
\begin{equation}
    \mathcal{R}_{up}^{U}[k] =\sum_{j =1}^{n \times (J \times Q)} \left | \mu_{k}\mu_{j} \psi^{\dagger^T}_{k} \psi^{\dagger}_{j}   \right |,
\end{equation}
and by $\mathcal{R}_{up}^{S}$ for the selected ones:
\begin{equation}
    \mathcal{R}_{up}^{S}[k] = \sigma^{2} \sum_{j=1}^{n \times (J \times Q)}   \left |  (\psi^{\dagger^T}_{k} \psi^{\dagger}_{j}) \psi_k^{T} \psi_j  \right |.
\end{equation}
we have that,
\begin{align}
    \mathcal{R}_{up}(x,W) & =  \sum_{k =1}^{n \times (J \times Q)} \min_{\delta_k}  \: \: 1_{\left \{ \delta_k =0 \right \}} \mathcal{R}_{up}^{U}[k]  + 1_{\left \{ \delta_k =1 \right \}} \mathcal{R}_{up}^{S}[k]  \\
    & =  \sum_{k =1}^{n \times (J \times Q)}  \min(\mathcal{R}_{up}^{U}[k],\mathcal{R}_{up}^{S}[k]).
\end{align}

\subsection{Comparison Upper Bound Ideal Risk with Orthogonal Ideal Risk}
\label{upper-bound-comparison}
\textbf{Proposition 1.}
\proof
The comparison of this upper bound risk given an orthogonal dictionary and the one derived in the orthogonal case is as follows:

 If the basis is orthogonal, we have,
\begin{equation}
 (\psi^{\dagger^T}_{k} \psi^{\dagger}_{j}) = \begin{cases} 1, & \mbox{if }k=j \\ 0, & \mbox{else}\end{cases}
\end{equation}
and,
\begin{equation}
    \psi_k^{T} \psi_j =  \begin{cases} 1, & \mbox{if }k=j \\ 0, & \mbox{else}\end{cases}
\end{equation}
Therefore, the upper-bound derived recovers the ideal risk in the orthogonal case.

\subsection{Comparison Upper Bound Ideal Risk with Empirical Risk}
\label{upper-bound-empirical}
\textbf{Proposition 2.}
\proof

If $D^S=I$, the empirical risk is equal to:
\begin{equation*}
   \tilde{\mathcal{R}}(y,W) = \sigma^{2} \sum_{k,j=1}^{n \times (J \times Q)}   \left |  (\psi^{\dagger^T}_{k} \psi^{\dagger}_{j}) \psi_k^{T} \psi_j  \right |.
\end{equation*}   
and the upper bound risk is:   
\begin{equation*} 
   \mathcal{R}_{up}(x,W) = \sigma^{2} \sum_{k,j=1}^{n \times (J \times Q)}   \left |  (\psi^{\dagger^T}_{k} \psi^{\dagger}_{j}) \psi_k^{T} \psi_j  \right |.
\end{equation*}
Thus both coincide as this restriction on the support of the risk makes it independent of both $x$ and $y$.
\textbf{Proposition 3.}
\proof
In the case where $D^U=I$,
\begin{align}
    \tilde{\mathcal{R}}(y,W) &= \sum_{k,j=1}^{n \times (J \times Q)} \left | \mu_k(y) \mu_j(y)  \psi^{\dagger^T}_{k}  \psi^{\dagger}_{j} \right | \\
    & = \sum_{k,j=1}^{n \times (J \times Q)}  \left | ( \mu_k(x) \mu_j(x) +  \mu_k(x) \mu_j(\epsilon) +  \mu_k(\epsilon) \mu_k(x) + \mu_k(\epsilon) \mu_j(\epsilon) ) \right | \times \nonumber \\ 
    & \quad \left | \psi^{\dagger^T}_{k}  \psi^{\dagger}_{j} \right |,
\end{align}
by the triangular inequality, we have that:
\begin{align}
    \tilde{\mathcal{R}}(y,W)\leq &  \sum_{k,j=1}^{n \times (J \times Q)} ( \left | \mu_k(x) \mu_j(x) \right | +\left |  \mu_k(x) \mu_j(\epsilon) \right | +  \left | \mu_k(\epsilon) \mu_k(x)\right | + \left | \mu_k(\epsilon) \mu_j(\epsilon) ) \right | \times \nonumber \\ 
    & \qquad \left |  \psi^{\dagger^T}_{k}  \psi^{\dagger}_{j} \right |,
\end{align}
by the monotony of expectation and the Fubini theorem, we have almost surely:

\begin{equation*}
   \tilde{\mathcal{R}}(y,W)\leq \mathcal{R}_{up}(x,W) +  \sum_{k,j=1}^{n \times (J \times Q)} C[k,j] \times \left |   \psi^{\dagger^T}_{k} \psi^{\dagger}_{j}   \right | \qquad a.s.,
\end{equation*}
where $C[k,j]$ is equals to,
\begin{equation*}
    C[k,j] =    \left | \mu_{k}(x) \right | \left \| \psi_j \right \|_1 \sigma \sqrt{\frac{2}{\pi}} +  \left | \mu_{j}(x) \right |\left \| \psi_k \right \|_1 \sigma \sqrt{\frac{2}{\pi}}  + \sigma^2 (1- \frac{2}{\pi}).
\end{equation*}
\end{document}